\documentclass[11pt, reqno]{amsart}
\usepackage{graphicx,amsmath,amssymb,epsfig}
\usepackage{rotating}

\usepackage{amsfonts}

\theoremstyle{plain}

\newtheorem{corollary}{Corollary}

\newtheorem{lemma}{Lemma}

\newtheorem{proposition}{Proposition}

\numberwithin{equation}{section}

\begin{document}
\title[ ]{Quantile and Probability Curves Without Crossing}
\author[ ]{Victor
Chernozhukov$^\dag$  \ \ Iv\'an Fern\'andez-Val$^\S$ \ \ Alfred
Galichon$^\ddag$   } \noindent

\date{This version of the paper is of
\today. Previous, more extended, versions (September 2006, April
2007) are available at www.mit.edu/$\sim$vchern/www and
www.ArXiv.org. The method developed in this paper has now been
incorporated in the package \texttt{quantreg} (Koenker, 2007) in
\texttt{R}. The title of this paper is (partially) borrowed from the
work of Xuming He (1997), to whom we are grateful for the
inspiration. We would like to thank
the editor Oliver Linton, three anonymous referees, Alberto Abadie,
Josh Angrist, Gilbert Bassett, Andrew Chesher, Phil Cross, James Durbin, Ivar
Ekeland, Brigham Frandsen, Raymond Guiteras, Xuming He, Roger
Koenker, Joonhwan Lee, Vadim Marmer, Ilya Molchanov, Francesca
Molinari, Whitney Newey, Steve Portnoy, Shinichi Sakata, Art
Shneyerov, Alp Simsek, and participants at  BU, CEMFI, CEMMAP
Measurement Matters Conference, Columbia Conference on Optimal
Transportation, Columbia, Cornell, Cowles Foundation 75th
Anniversary Conference, Duke-Triangle, Ecole Polytechnique,
Frontiers of Microeconometrics in Tokyo, Georgetown, Harvard-MIT,
MIT, Northwestern, UBC, UCL,  UIUC, University of Alicante, and
University of Gothenburg Conference ``Nonsmooth Inference,
 Analysis, and Dependence," for comments that
helped us to considerably improve the paper.  We are grateful to
Alberto Abadie for providing us the data for the empirical example.
The authors gratefully acknowledge research support from the
National Science Foundation and chaire X-Dauphine ``Finance et
D\'eveloppement Durable''.}

\thanks{$\dag$ Massachusetts Institute of Technology, Department of
Economics and Operations Research Center, University College London,
CEMMAP. E-mail: vchern@mit.edu.}

\thanks{$\S$ Boston University, Department of Economics. E-mail:
ivanf@bu.edu.}

\thanks{$\ddag$ Ecole Polytechnique, D$\acute{e}$partement
d'Economie. E-mail: alfred.galichon@polytechnique.edu.}

 \maketitle

\begin{abstract}

This paper proposes a method to address the longstanding problem of
lack of monotonicity in estimation of conditional and structural
quantile functions, also known as the quantile crossing problem (Bassett and Koenker, 1982). The
method consists in sorting or monotone rearranging the original
estimated non-monotone curve into a monotone rearranged curve. We
show that the rearranged curve is closer to the true quantile curve
in finite samples than the original curve, establish a functional
delta method for rearrangement-related operators, and derive
functional limit theory for the entire rearranged curve and its
functionals. We also establish validity of the bootstrap for
estimating the limit law of the the entire rearranged curve and its
functionals. Our limit results are generic in that they apply to
every estimator of a monotone econometric function, provided that
the estimator satisfies a functional central limit theorem and the
function satisfies some smoothness conditions. Consequently, our
results apply to estimation of other econometric functions with
monotonicity restrictions, such as demand, production, distribution,
and structural distribution functions. We illustrate the results
with an
application to estimation of structural distribution and quantile functions using data on Vietnam veteran status and earnings. \\



\end{abstract}

\newpage

\section{Introduction}

This paper addresses the longstanding problem of lack of
monotonicity in the estimation of conditional and structural
quantile functions, also known as the quantile crossing problem
(Bassett and Koenker, 1982, and He, 1997). The most common approach
to estimating quantile curves is to fit a curve, often linear,
pointwise for each probability index.\footnote{This includes all
principal approaches to estimation of conditional quantile
functions, such as the canonical quantile regression of Koenker and
Bassett (1978) and censored quantile regression of Powell (1986).
This also includes principal approaches to estimation of structural
quantile functions, such as the instrumental quantile regression
methods via control functions of Imbens and Newey (2001), Blundell
and Powell (2003), Chesher (2003), and Koenker and Ma (2006), and
instrumental quantile regression estimators of Chernozhukov and
Hansen (2005, 2006).} Researchers use this approach for a number of
reasons, including parsimony of the resulting approximations and
excellent computational properties. The resulting fits, however, may
not respect a logical monotonicity requirement -- that the quantile
curve should be increasing as a function of the probability index.

This paper introduces a natural monotonization of the empirical
curves by sampling from the estimated non-monotone model, and then
taking the resulting conditional quantile curves which by
construction are monotone in the probability index. This
construction of the monotone curve may be seen as a bootstrap and as
a sorting or monotone rearrangement of the original non-monotone
curve (see Hardy et al., 1952, and references given below). We show
that the rearranged curve is closer to the true quantile curve in
finite samples than the original curve is, and derive functional
limit distribution theory for the rearranged curve to perform
simultaneous inference on the entire quantile function. Our theory
applies to both dependent and independent data, and to a wide
variety of original estimators, with only the requirement that they
satisfy a functional central limit theorem. Our results also apply
to many other econometric problems with monotonicity restrictions,
such as distribution and structural distribution functions, as well
as demand and production functions, option pricing functions, and
yield curves.\footnote{See Matzkin (1994) for more examples and
additional references, and Chernozhukov et. al. (2009) for further
theoretical results that cover the latter set of applications. } As
an example, we provide an empirical application to estimation of
structural distribution and quantile functions based on Abadie
(2002) and Chernozhukov and Hansen (2005, 2006).

There exist other methods to obtain monotonic fits for conditional
quantile functions. He (1997), for example, proposed to impose a
location-scale regression model, which naturally satisfies
monotonicity. This approach is fruitful for location-scale
situations, but in numerous cases the data do not satisfy the
location-scale paradigm, as discussed in Lehmann (1974), Doksum
(1974), and Koenker (2005). Koenker and Ng (2005) developed a
computational method for quantile regression that imposes the
non-crossing constraints in simultaneous fitting of a finite number
of quantile curves. The statistical properties of this method have
yet to be studied, and the method does not immediately apply to
other quantile estimation methods.  Mammen (1991) proposed two-step
estimators, with mean estimation in the first step followed by
isotonization in the second.\footnote{Isotonization is also known as
the ``pool-adjacent-violators algorithm" in statistics and
``ironing" in economics. It amounts to projecting the original
estimate on the set of monotone functions.} Similarly to Mammen
(1991), we can employ quantile estimation in the first step followed
by isotonization in the second, obtaining an interesting method
whose properties have yet to be studied.  In contrast, our method
uses rearrangement rather than isotonization, and is better suited
for quantile applications. The reason is that isotonization is best
suited for applications with (near) flat target functions, while
rearrangement is best suited for applications with steep target
functions, as in typical quantile applications. Indeed, in a
numerical example closely matching our empirical application, we
find that rearrangement significantly outperforms isotonization.
Finally, in an independent and contemporaneous work, Dette and
Volgushev (2008) propose to obtain monotonic quantile curves by
applying an integral transform to a local polynomial estimate of the
conditional distribution function, and derive pointwise limit theory
for this estimator. In contrast, we directly monotonize any generic
estimate of a conditional quantile function and then derive generic
functional limit theory for the entire monotonized
curve.\footnote{We refer to Dette and Volgushev (2008) for a more
detailed comparison of the two approaches.}

In addition to resolving the problem of estimating quantile curves
that avoid crossing, this paper develops a number of original
theoretical results on rearranged estimators. It therefore makes
both practical and theoretical contributions to econometrics and
statistics.   In order to discuss these contributions more
specifically, it is helpful first to review some of the relevant
literature and available results.

We begin the review by noting that the idea of rearrangement goes
back at least to Chebyshev (see Bronstein et al., 2003, p. 31, Hardy
et al., 1952, and Lorentz, 1953, among others). Rearrangements have
been extensively used in functional analysis and operations research
(Villani, 2003, and Carlier and Dana, 2005), but not in econometrics
or statistics until recently.   Recent research on rearrangements in
statistics include the work of Fougeres (1997), which used
rearrangement to produce a monotonic kernel density estimator and
derived its uniform rates of convergence; Davydov and Zitikis
(2005), which considered tests of monotonicity based on rearranged
kernel mean regression; Dette et al. (2006) and Dette and Scheder
(2006), which introduced smoothed rearrangements for kernel mean
regressions and derived pointwise limit theory for these estimators;
and Chernozhukov et al. (2009), which used univariate and
multivariate rearrangements on point and interval estimators of
monotone functions based on series and kernel regression estimators.
In the context of our problem, rearrangement is also connected to
the quantile regression bootstrap of Koenker (1994). In fact, our
research grew from the realization that we could use this bootstrap
for the purpose of monotonizing quantile regressions, and we
discovered the link to the classical procedure of rearrangement
later, while reading Villani (2003).

The theoretical contributions of this paper are threefold. First,
our paper derives functional limit theory for rearranged estimators
and functional delta methods for rearrangement operators, both of
which are important original results. Second, the paper derives
functional limit theory for estimators obtained by
rearrangement-related operations, which are also original results.
For example, our theory includes as a special case the asymptotics
of the conditional distribution function estimator based on quantile
regression, whose properties have long remained unknown (Bassett and Koenker, 1982). Moreover,
our limit theory applies to functions,  encompassing the pointwise
results as a special case. An attractive feature of our theoretical results is that
they do \textit{not} rely on independence of data, the particular
estimation method used, or any parametric assumptions. They only
require that a functional central limit theorem applies to the
original estimator of the curve, and the population curves have some
smoothness properties. Our results therefore apply to any quantile
model and quantile estimator that satisfy these requirements. Third,
our results immediately yield validity of the bootstrap for
rearranged estimators, which is an important result for practice.

We organize the rest of the paper as follows.  In Section 2 we
present some analytical results on rearrangement and then present
all the main results; in Section 3 we provide an application and a
numerical experiment that closely matches the application; in
Section 4 we give some concluding remarks; and in the Appendix we
include the proofs of the results. The data and programs used for
the examples are available in the on-line supplement (Chernozhukov
et al., 20??).

\section{Rearrangement: Analytical and Empirical Properties}\label{section:
rearrangement}

In this section, we describe rearrangement, derive some basic
analytical properties of the rearranged curves in the population,
establish functional differentiability results, and establish
functional limit theorems and other estimation properties.

\subsection{Rearrangement}  We consider a target function $u \mapsto Q_0(u|x)$ that, for each
$x \in \mathcal{X}$, maps $[0,1]$  to the real line and is increasing
 in $u$.  Suppose that $u \mapsto \widehat{Q}(u|x)$ is a
parametric or nonparametric estimator of $Q_0(u|x)$.   Throughout
the paper, we use conditional and structural quantile estimation as
the main application, where $u \mapsto Q_0(u|x)$ is the quantile
function of a real response variable $Y$, given a vector of
regressors $X=x$.  Accordingly, we will usually refer to the
functions $u \mapsto Q_0(u|x)$ as quantile functions throughout the
paper. In other applications, such as estimation of conditional and
structural distribution functions, other names would be appropriate
and we need to accommodate different domains, as described in Remark
1 below.

Typical estimation methods fit the quantile function
$\widehat{Q}(u|x)$ pointwise in $ u \in [0,1]$.\footnote{See Koenker
and Bassett (1978), Powell (1986), Chaudhuri (1991), Buchinsky and
Hahn (1998), Yu and Jones (1998), Abadie et al. (2002), Honor\'e et
al. (2002), and Chernozhukov and Hansen (2006),  among others, for
examples of exogenous, censored, endogenous, nonparametric,  and
other types of quantile regression estimators.} A problem that might
occur is that the map $ u \mapsto \widehat Q(u|x)$ may not be
increasing in $u$, which violates the logical monotonicity
requirement. Another manifestation of this issue, known as the
quantile crossing problem, is that the conditional quantile curves
$x \mapsto \widehat Q(u|x)$ may cross for different values of $u$
(He, 1997).  Similar issues also arise in estimation of conditional
and structural distribution functions (Hall et al., 1999, and
Abadie, 2002).

We can transform the possibly non-monotone  function $ u \mapsto
\widehat Q(u|x)$ into a monotone function  $u \mapsto \widehat
Q^*(u|x)$
by quantile bootstrap or rearrangement. That is, we consider the random variable $Y_{x} := \widehat Q(U|x)$ where $%
U\sim \text{Uniform}(\mathcal{U})$ with $\mathcal{U}=[0,1]$, and
take its quantile function denoted by $u \mapsto \widehat Q^*(u|x)$
instead of the original function $ u \mapsto \widehat Q(u|x)$. This
variable $Y_x$ has a distribution function:
\begin{equation}\label{stepa}
\widehat{F}(y|x) := \int_{0}^{1}1\{\widehat Q(u|x)\leq y\}d u ,
\end{equation}
which is naturally monotone in the level $y$, and a quantile
function:
\begin{equation}\label{stepb}
\widehat{Q}^{\ast}(u | x) :=  \widehat{F}^{-1}( u |x)=\inf
\{y:\widehat{F}(y|x)\geq u \},
\end{equation}
which is naturally monotone in the index $u$.  Thus, starting with a
 possibly non-monotone original curve $ u \mapsto \widehat Q(u|x)$, the rearrangement
(\ref{stepa})-(\ref{stepb}) produces a monotone quantile curve $u
\mapsto \widehat{Q}^{\ast}( u |x)$. Of course, the rearranged
quantile function $u \mapsto \widehat{Q}^{\ast}(u|x)$ coincides with
the original function $u \mapsto \widehat Q(u|x)$ if the original
function is non-decreasing
 in $u$, but differs from it otherwise.

The mechanism (\ref{stepa})-(\ref{stepb}) and its name have a direct
relation to the rearrangement operator from functional analysis
(Hardy et al., 1952), since $u \mapsto \widehat{Q}^{\ast}( u |x)$ is
the monotone rearrangement of $u \mapsto \widehat Q(u|x)$.
Equivalently, as we stated earlier, rearrangement has a direct
relation to the quantile bootstrap (Koenker, 1994), since the
rearranged quantile curve is the quantile function of the bootstrap
variable produced by the estimated quantile model. Moreover, we
refer the reader to Dette et al. (2006, p. 470) who, using a closely
related motivation, introduced the idea of smoothed rearrangement,
which produces smoothed versions of (\ref{stepa}) and (\ref{stepb})
and can be valuable in applications. Finally, for practical and
computational purposes, it is helpful to think of rearrangement as
sorting. Indeed to compute the rearrangement of a continuous
function $u \mapsto \widehat Q(u|x)$  we simply set $\widehat
Q^\ast(u|x)$ as the u-th quantile of $ \{ \widehat Q(u_1|x),...,
\widehat Q(u_k|x) \}$, where $\{u_1,...,u_k\}$ is a sufficiently fine
net of equidistant indices in $[0,1]$.

\textbf{Remark 1.}(\emph{Adjusting for domains different from the
unit interval}). Throughout the paper we assume that the domain of
all the functions is the unit interval, $\mathcal{U}=[0,1]$, but in
many applications we may have to deal with different domains. For
example, in quantile estimation problems, we may consider a
subinterval $[a,b]$ of the unit interval as the domain, in order to
avoid estimation of tail quantiles.  In distribution estimation
problems, we may consider the entire real line as the domain.  In
such cases we can first transform these functions to have the unit
interval as  the domain. Concretely, suppose we have an original
function $\bar Q: [a, b] \to \Bbb{R}$. Then using any increasing
bijective mapping $\varphi :[a, b]  \mapsto [0,1]$, we can define $
Q := \bar Q \circ \varphi^{-1}: [0,1] \to \Bbb{R}$, and then proceed
to obtain its rearrangement $Q^*$.  In the case where $ a \neq
-\infty $ and $b \neq \-\infty$, we can take $\varphi$ to be an
affine mapping. In order to obtain an increasing rearrangement $\bar
Q^*$ of $\bar Q$, we then set $ \bar Q^* = Q^* \circ \varphi$. \qed

Let $Q$ denote the pointwise probability limit of $\widehat{Q}$, which we
will refer to as the population curve. In the analysis we distinguish the
following two cases:
\begin{enumerate}
\item Monotonic  $Q$: The population curve $ u \mapsto Q(u|x)$ is
    increasing in $u$, and thus satisfies the monotonicity requirement.
\item  Non-monotonic $Q$: The population curve $ u \mapsto Q(u|x)$ is
    non-monotone due to misspecification.
    \end{enumerate}
In case (1) the empirical curve $ u \mapsto
    \widehat Q(u|x)$ may be non-monotone due to estimation error, while in case (2) it may be
non-monotone due to both misspecification and estimation error.  A
leading example of case (1) is  when the population curve $Q$ is
correctly specified, so that it equals the target quantile curve,
namely $Q(u|x) = Q_0(u|x)$ for all $ u \in [0,1]$. Case (1) also
allows for some degree of misspecification, provided that the
population curve, $Q \neq Q_0,$ remains monotone.  A leading example
of  case (2) is   when the population curve $Q$ is misspecified, $Q
\neq Q_0$, to a degree that makes $u \mapsto Q(u|x)$ non-monotone.
For example, the common linear specification $u \mapsto Q(u|x) =
p(x)^\mathsf{T}\beta(u)$ can be non-monotone if the support of $X$
is sufficiently rich, while the set of transformations of $x$,
$p(x),$ is not (Koenker, 2005, Chap 2.5). Typically, by using a rich
enough set $p(x)$  we can approximate the true function  $Q_0(u|x)$
sufficiently well, and thus often avoid case (2), see Koenker, 2005,
Chap 2.5.  This is the strategy that we generally recommend, since
inference and limit theory under case (1) is theoretically and
practically simpler than under case (2).  However, in what follows
we analyze the behavior of rearranged estimators both in cases (1)
and (2), since either of these cases could occur.



In the rest of the section, we establish the empirical properties of the
rearranged estimated quantile functions and the corresponding distribution
functions:
\begin{equation} \label{eq: empirical curves}
 u \mapsto \widehat{Q}^{\ast}(u|x) \ \text{ and } \  y \mapsto \widehat
 F(y|x),
\end{equation}
under cases (1) and (2).


\subsection{Basic Analytical Properties of Population Curves}
We start by characterizing certain analytical properties of the
probability limits or population versions  of empirical curves
(\ref{eq: empirical curves}), namely
\begin{equation}\label{eq: define F and Q}
\begin{array}{ll}
 y \mapsto F(y|x) =
\int_{0}^{1}1\{Q(u|x) \leq y\}d u, \\
u \mapsto  Q^{\ast}(u|x) := F^{-1}(u|x) =\inf \{y: F(y|x)\geq  u \}.
\end{array}
\end{equation}
We need these properties to derive our main limit results stated in the
following subsections.

Recall first the following definitions from Milnor (1965). Let $g:
\mathcal{U} \subset \mathbb{R} \mapsto \mathbb{R} $ be a
continuously differentiable function. A point $u\in \mathcal{U}  $
is called a regular point of $g$ if the derivative of $g$ at this
point does not vanish, i.e., $\partial_u g\left( u \right) \neq 0$,
where $\partial_u$ denotes the partial derivative operator with
respect to $u$ . A point $u$ which is not a regular point is called
a critical point. A value $y\in g\left( \mathcal{U}  \right) $ is
called a regular value of $g$ if $g^{-1}( y ) $ contains only
regular points, i.e., if $\forall u \in g^{-1} ( y ) $, $\partial_u
g \left( u \right) \neq 0$. A value $y$ which is not a regular value
is called a critical value.

Define region $\mathcal{Y}_{x}$ as the support of $Y_{x}$, and
regions $\mathcal{YX} :=\left\{ \left( y,x \right) :y \in
\mathcal{Y}_{x}, x \in \mathcal{X} \right\} $ and $\mathcal{UX}
:=\mathcal{U}\times\mathcal{X}$. We assume throughout that
$\mathcal{Y}_x \subset \mathcal{Y}$, a compact subset of $\Bbb{R}$,
and that $x \in \mathcal{X}$, a compact subset of $\Bbb{R}^d$. In
some applications the curves of interest are not functions of $x$,
or we might be interested in a particular value $x$. In this case,
we can take the set $\mathcal{X}$ to be a singleton
$\mathcal{X}=\{x\}$.

\noindent \textbf{Assumption 1.} (Properties of $Q$). \textit{We
maintain the following assumptions on $Q$ throughout the paper:}
\begin{enumerate}
\item[(a)]  \textit{$Q: \mathcal{U}\times \mathcal{X} \mapsto
    \mathbb{R}$ is a continuously differentiable function in both
    arguments.}
\item[(b)]  \textit{The number of elements of
    $\{u \in \mathcal{U} \ | \ \partial_u Q(u|x)=0\}$ is uniformly
    bounded on $x \in \mathcal{X}$.}
 \end{enumerate}

Assumption 1(b) implies that, for each $x \in \mathcal{X}$,
$\partial_u Q(u|x)$ is not zero almost everywhere on $\mathcal{U}$
and can switch sign only a bounded number of times. Further, we
define $\mathcal{Y}_{x}^{\ast } $ be the subset of regular values of
$u \mapsto Q( u |x)$ in $\mathcal{Y}_{x}$, and $\mathcal{YX}^{\ast
}:=\left\{ \left( y,x \right) :y \in \mathcal{Y}_{x}^{\ast}, x \in
\mathcal{X} \right\} $.

We use a simple example to describe some basic analytical properties
of (\ref{eq: define F and Q}), which we state more formally in the
proposition given below. Consider the following pseudo-quantile
function: $Q(u) = 5 \{ u + \sin(2 \pi u)/\pi \},$ which is highly
non-monotone in $[0,1]$ and therefore fails to be a proper quantile
function. The left panel of Figure 1 shows $Q$ together with its
monotone rearrangement $Q^*$. We see that $Q^*$ partially coincides
with $Q$ on the areas where $Q$ behaves like a proper quantile
function, and that $Q^*$ is continuous and increasing.  Note also
that $1/3$ and $2/3$ are the critical points of $Q$, and $3.04$ and
$1.96$ are the corresponding critical values. The right panel of
Figure 1 shows the pseudo-distribution function $Q^{-1}$, which is
multi-valued, and the distribution function $F={Q^*}^{-1}$ induced
by sampling from $Q$.  We see that $F$ is continuous and does not
have point masses. The left panel of Figure 2 shows $\partial_u
Q^*$, the sparsity function for $Q^*$.  We see that the sparsity
function is continuous at the ${Q^*}^{-1}$-image of the regular
values of $Q$ and has jumps at the ${Q^*}^{-1}$-image of the
critical values of $Q$.    The right panel of Figure 2 shows
$\partial_y F$, the density function for $F$. We see that
$\partial_y F$ is continuous at the regular values of $Q$ and has
jumps at the critical values of $Q$.

\begin{figure}
\begin{center}
\epsfig{figure=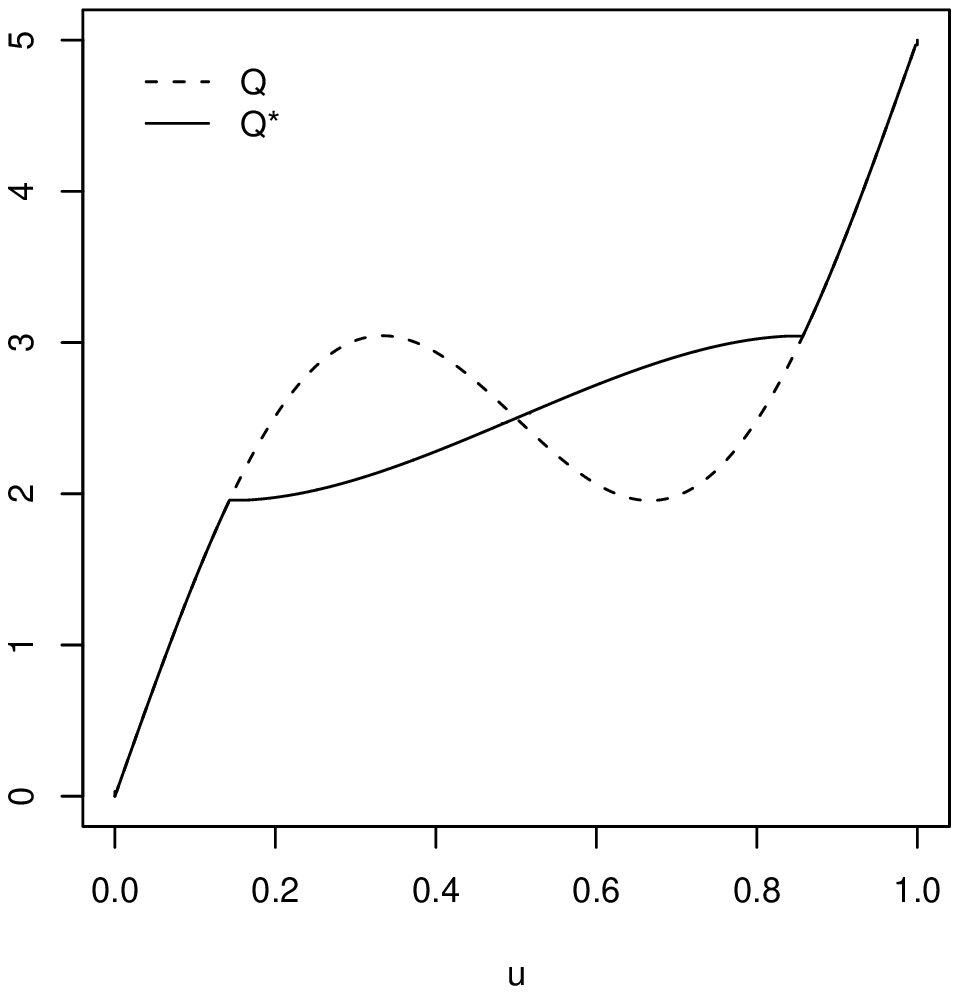,
width=3in,height=3.2in}\epsfig{figure=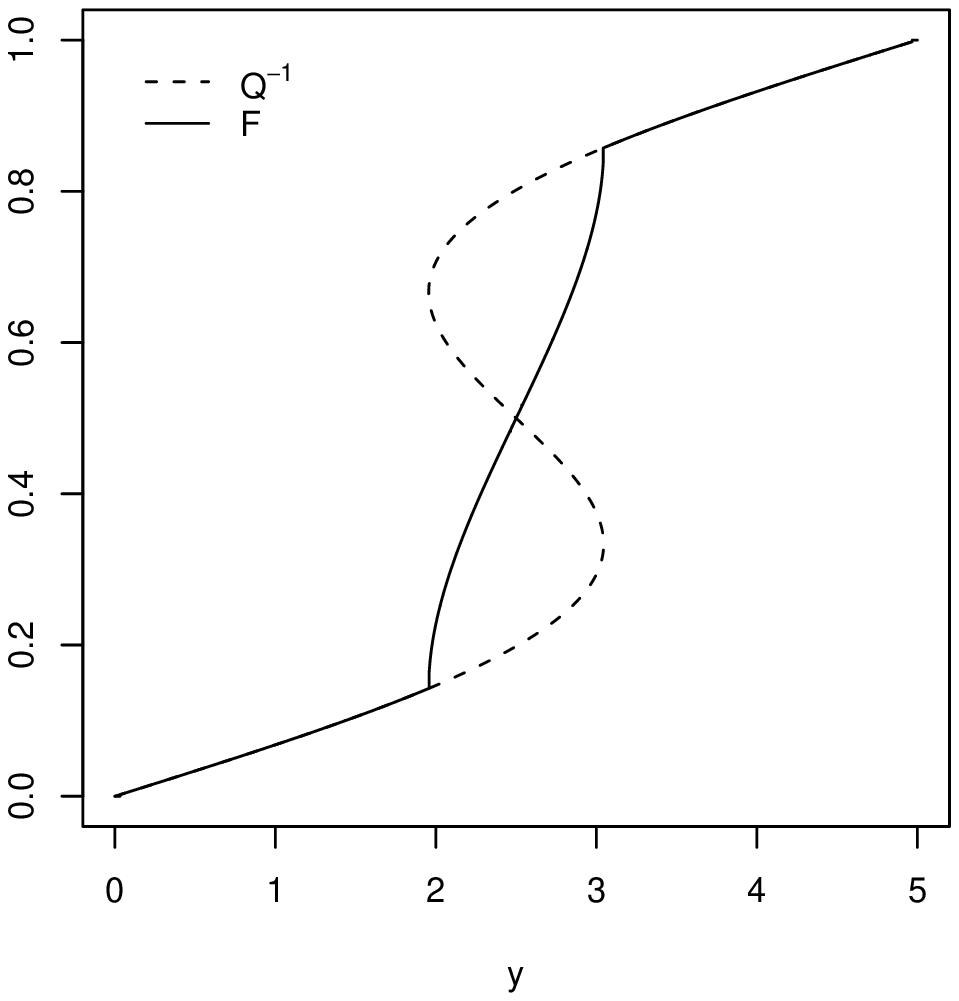,
width=3in,height=3.2in}\caption{\label{Fig:1} Left: The
pseudo-quantile function $Q$ and the rearranged quantile function
$Q^{\ast}$. Right: The pseudo-distribution function $Q^{-1}$ and the
distribution function $F$ induced by $Q$.}
\end{center}
\begin{center}
\epsfig{figure=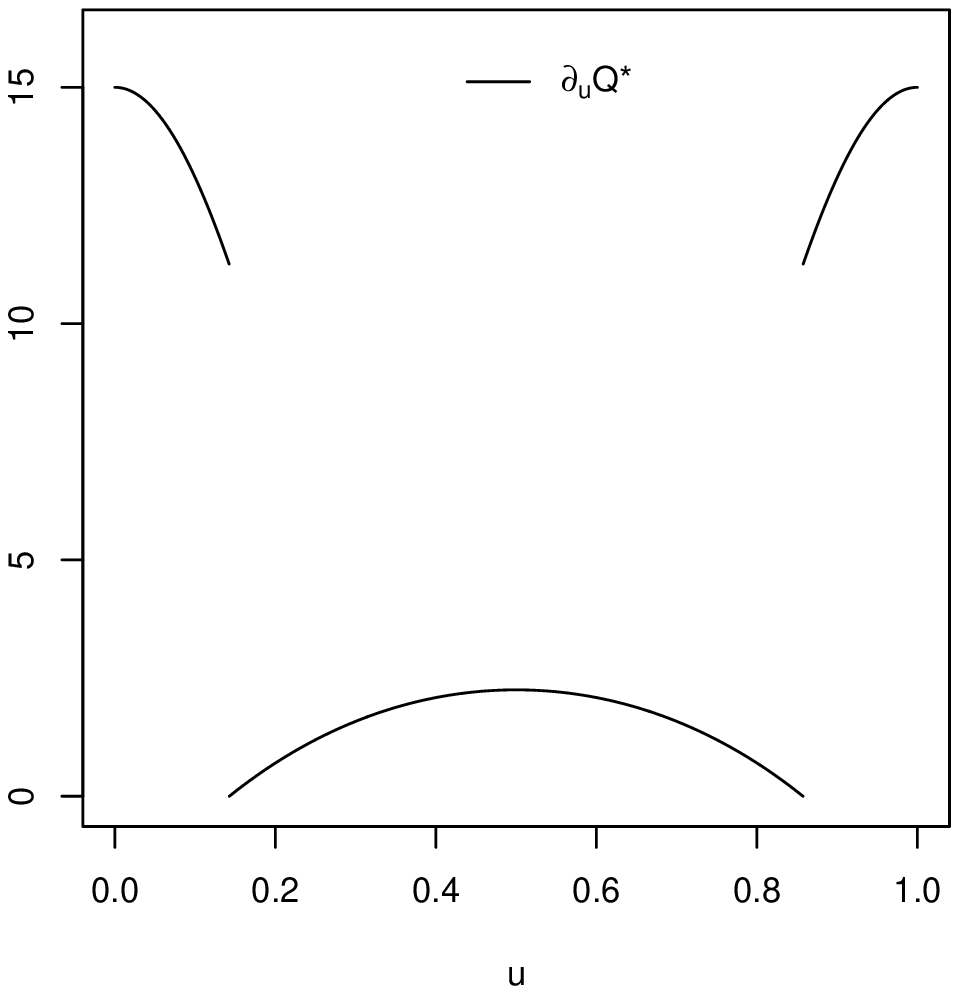,
width=3in,height=3.2in}\epsfig{figure=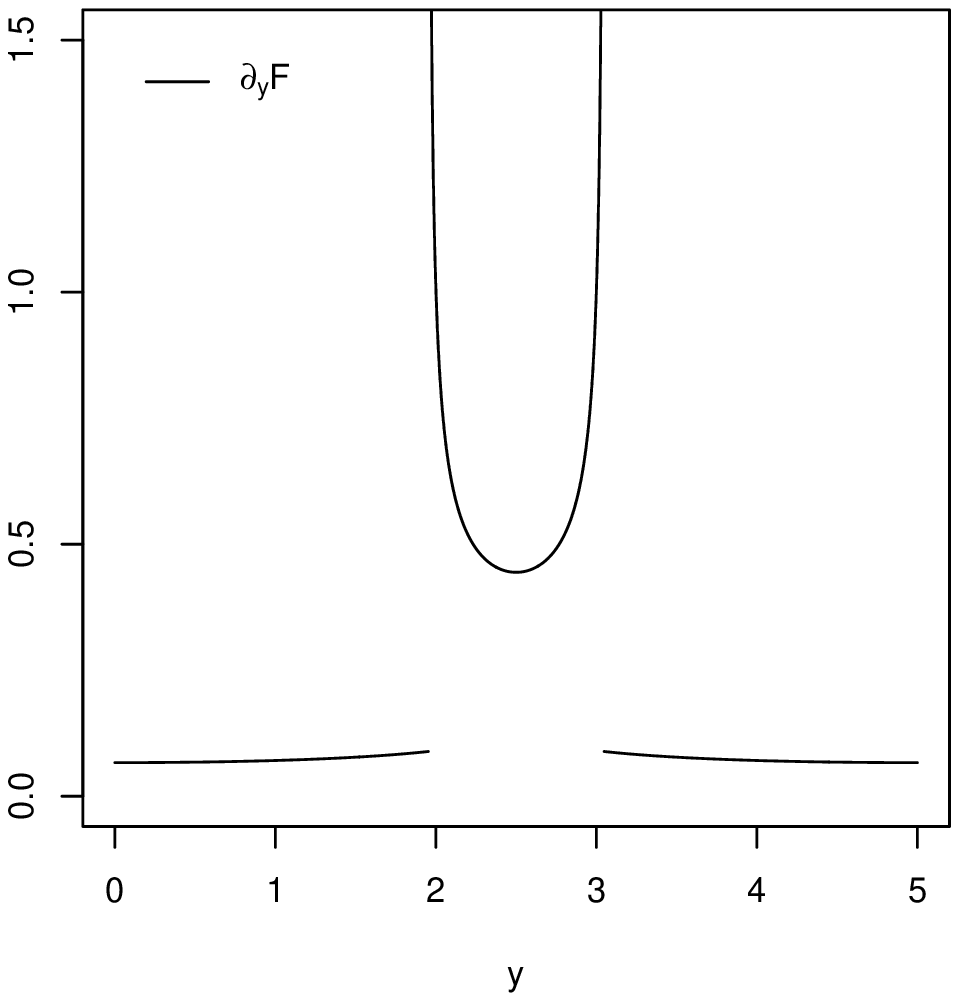,
width=3in,height=3.2in}\caption{\label{Fig:2} Left: The density
(sparsity) function of the rearranged quantile function $Q^{\ast}$.
Right: The density function of the distribution function $F$ induced
by $Q$.
 }
\end{center}
\end{figure}

The following proposition states more formally the properties of
$Q^*$ and $F$:
\begin{proposition}[Basic properties of $F$ and $Q^{\ast}$]
The functions $y \mapsto F(y|x)$ and $u \mapsto Q^{\ast}(u|x)$
satisfy the following properties, for each $x \in \mathcal{X}$: (1) The set of critical values, $\mathcal{Y}_{x}\setminus \mathcal{Y}%
_{x}^{\ast }$, is finite, and $\int_{\mathcal{Y}_{x}\setminus \mathcal{Y}%
_{x}^{\ast }}dF(y|x)=0$. (2) For any $y\in \mathcal{Y}_{x}^{\ast },$
\begin{equation*}
F(y|x)=\sum_{k=1}^{K(y|x)}\text{sign}%
\{\partial_u Q( u _{k}(y|x)|x)\} u _{k}(y|x)+1\{\partial_u Q( u
_{K(y|x)}(y|x)|x)<0\},
\end{equation*}
where $\{ u _{k}(y|x),\text{ for }k=1,2, ...,K(y|x)<\infty \}$ are
the
roots of $Q( u |x)=y$ in increasing order. (3) For any $y\in \mathcal{Y}^{\ast }_{x}$, the ordinary derivative $%
f(y|x):=\partial_y F(y|x)$ exists and takes the form
\begin{equation*}
f(y|x)=\sum_{k=1}^{K(y|x)}\frac{1}{|\partial_u Q( u _{k}(y|x)|x)|},
\end{equation*}%
which is continuous at each $y\in \mathcal{Y}_{x}^{\ast }$. For any
$y \in \mathcal{Y} \setminus \mathcal{Y}_{x}^{\ast }$, we set
$f(y|x) := 0$. $F(y|x)$ is
absolutely continuous and strictly increasing in $%
y\in \mathcal{Y}_{x}$. Moreover, $y \mapsto f(y|x)$ is a Radon-Nikodym derivative of $%
y \mapsto F(y|x)$ with respect to the Lebesgue measure. (4) The quantile function $u \mapsto Q^{\ast}(u|x)$ partially coincides with $%
u \mapsto Q(u|x)$; namely $Q^{\ast}( u |x) = Q( u |x),$ provided
that $u \mapsto Q(u|x)$ is increasing at $u$, and the preimage of $
Q^{\ast}(u|x)$ under $Q$ is unique.
(5) The
quantile function $u \mapsto Q^{\ast}(u|x)$ is equivariant to
monotone transformations of $u \mapsto Q(u|x)$, in particular, to
location and scale transformations. (6) The quantile function $u
\mapsto Q^{\ast}(u|x)$ has an ordinary continuous derivative $
\partial_u Q^{\ast}( u |x) = 1/f(Q^{\ast}( u |x)|x),$ when
$Q^{\ast}( u |x)\in \mathcal{Y}_{x}^{\ast }$. This function is also
a Radon-Nikodym derivative with respect to the Lebesgue measure.
(7)  The map $(y,x) \mapsto F(y|x)$ is continuous on $\mathcal{YX}$
and the map $(u,x) \mapsto Q^{\ast}(u|x)$ is continuous on
$\mathcal{UX}$.

\end{proposition}

\subsection{Functional Derivatives for Rearrangement-Related Operators}

Here we derive  functional derivatives for the rearrangement
operator $Q \mapsto Q^*$ and the pre-rearrangement operator
$Q\mapsto F$ defined by equation (\ref{eq: define F and Q}).  These
results constitute the first set of original main theoretical
results obtained in this paper.    In the subsequent sections, these
results allow us to establish a generic functional central limit
theorem for the estimated functions $\widehat Q^\ast$ and $\widehat
F$, as well as to establish validity of the bootstrap for estimating
their limit laws.

In order to describe the results, let $\ell^{\infty} (\mathcal{U}
\mathcal{X})$ denote the set of bounded and measurable functions $h:
\mathcal{U} \mathcal{X} \mapsto \Bbb{R}$, $C(\mathcal{U}
\mathcal{X})$ the set of continuous functions $h: \mathcal{U}
\mathcal{X} \mapsto \Bbb{R}$, and $\ell^1(\mathcal{U} \mathcal{X})$
the set of measurable functions $h: \mathcal{U} \mathcal{X} \mapsto
\Bbb{R}$ such that $\int_{\mathcal{U}}\int_{\mathcal{X}} |h(u|x)| du
dx $ $ < \infty$, where $du$ and $dx$ denote the integration with
respect to the Lebesgue measure on $\mathcal{U}$ and $\mathcal{X}$,
respectively.

\begin{proposition} [Hadamard derivatives of $F$ and $Q^{\ast}$ with respect to
$Q$] \label{proposition: hadamard_diff}  (1) Define
$F(y|x,h_{t}):=\int_{0}^{1} 1 \{Q(u|x )+th_{t}( u |x)\leq y\}d u $.
As $t\rightarrow 0$,
\begin{eqnarray}\label{D1}
& & D_{h_t}(y|x,t) := \frac{F(y|x,h_{t})-F(y|x)}{t}\rightarrow
D_{h}(y|x), \\
& & D_{h}(y|x):=-\sum_{k=1}^{K(y|x)}\frac{h( u _{k}\left( y|x\right) |x)}{%
|\partial_u Q( u _{k}(y|x)|x)|}. \label{D1b}
\end{eqnarray}%
The convergence holds uniformly in any compact subset of
$\mathcal{YX}^{\ast}:=\{(y,x): y \in \mathcal{Y}_x^{\ast}, x\in \mathcal{X}\} $,
for every $|h_{t}-h|_{\infty}\rightarrow 0$, where $%
h_{t} \in \ell^{\infty} \left(\mathcal{U} \mathcal{X} \right)$, and
$h \in C(\mathcal{U} \mathcal{X})$.  (2) Define $Q^\ast(u|x,h_{t}):=
F^{-1}(y|x,h_{t}) = \inf\{y : F(y|x,h_{t}) \geq u\}$.
 As $t\rightarrow 0$,
\begin{eqnarray}
& & \tilde D_{h_t}(u|x,t):=\frac{Q^{\ast}( u |x,h_{t})-Q^{\ast}( u
|x)}{t}\rightarrow \tilde D_{h}(u|x),
\\ & & \tilde D_{h}(u|x) := -\frac{1}{%
f(Q^{\ast}( u |x)|x)} \  D_{h}(Q^{\ast}( u |x)|x).\label{D2b}
\end{eqnarray}%
The convergence holds uniformly in any compact subset of
$\mathcal{UX}^*= \{( u,x) : (Q^{\ast}( u |x),x)\in
\mathcal{YX}^*\}$, for every $|h_{t}-h|_{\infty}\rightarrow 0$, where $%
h_{t} \in \ell^{\infty} \left( \mathcal{U} \mathcal{X} \right)$, and
$h \in C(\mathcal{U} \mathcal{X})$.
\end{proposition}

This proposition establishes the Hadamard (compact)
differentiability of the rearrangement operator $Q \mapsto Q^*$ and
the pre-rearrangement operator $Q \mapsto F$ with respect to $Q$,
tangentially to the subspace of continuous functions.  Note that the
convergence holds uniformly on regions that exclude the critical
values of the mapping $u \mapsto Q(u|x)$. These results are new and
could be of independent interest. Rearrangement operators include
inverse (quantile) operators as a special case. In this sense, our
results generalize the previous results of  Gill and Johansen
(1990), Doss and Gill (1992), and Dudley and Norvaisa (1999) on
functional delta method (Hadamard differentiability) for the
quantile operator.   There are two main difficulties in establishing
the Hadamard differentiability in our case: first, like in the
quantile case,  we allow the perturbations $h_t$ to $Q$ to be
discontinuous functions, though converging to continuous functions;
second, unlike in the quantile case, we allow the perturbed
functions $Q + t h_t$ to be non-monotone even when $Q$ is monotone.
We need to allow for such rich perturbations in order to match
applications where the empirical perturbations $h_t = (\widehat Q -
Q)/t,$ for $t = 1/a_n$ and $a_n$ a growing sequence with the sample
size $n,$ are discontinuous functions, though converging to
continuous functions by the means of a functional central limit
theorem; moreover, the empirical (pseudo) quantile functions
$\widehat Q = Q + t h_t$ are not monotone even when $Q$ is monotone.

The following result deals with the monotonic case. It is worth
emphasizing separately, because functional derivatives are
particularly simple and we do not have to exclude any non-regular
regions from the domains.

\begin{corollary}[Hadamard derivatives of $F$ and $Q^{\ast}$ with respect to
$Q$ in the monotonic case]  Suppose $u\mapsto Q(u|x)$ has
$\partial_u Q(u|x)> 0$, for each $(u,x)\in \mathcal{UX}$. Then
$\mathcal{YX}^* = \mathcal{YX}$ and $\mathcal{UX}^* = \mathcal{UX}$.
Therefore, the convergence in Proposition \ref{proposition:
hadamard_diff} holds uniformly over the entire $\mathcal{YX}$ and
$\mathcal{UX}$, respectively. Moreover, $\tilde D_{h}(u|x) = h$,
i.e., the Hadamard derivative of the rearranged function with
respect to the original function is the identity operator.
\end{corollary}

Next we consider the following linear functionals obtained by
integration:
\begin{equation*}\begin{split}
&  (y',x)\mapsto \int_{\mathcal{Y}} g(y|x,y') F(y|x) dy, \ \
(u',x)\mapsto  \int_{\mathcal{U}} g(u|x, u') Q^{\ast}(u|x) d u,
\end{split}\end{equation*}
with the restrictions on $g$ specified below. These functionals are
of interest because they are useful building blocks for various
statistics, for example, Lorenz curves with function $g(u|x, u')= 1
\{ u \leq u'\}$, as discussed in the next section. The following
proposition calculates the Hadamard derivative of these functionals.

\begin{proposition}[Hadamard derivative of linear functionals of $Q^\ast$ and $F$
with respect to $Q$] \label{proposition: hadamard_diff linear_funct}
The following results are true with the limits being continuous on
the specified domains:
\begin{equation*}
1. \quad  \int_{\mathcal{Y}}g(y|x,y') D_{h_t}(y|x,t) d
y \to  \int_{\mathcal{Y}}  g(y|x, y') D_{h}(y|x)    dy
\end{equation*}
uniformly in $(y',x) \in \mathcal{Y} \mathcal{X}$, for any
 measurable $g$ that is bounded uniformly in its arguments and
 such that $(x,y') \mapsto g(y|x, y')$ is
 continuous  for a.e. $y$.
\begin{equation}\label{equation: had_diff_lin_fun_part2}
2. \quad   \int_{\mathcal{U}} g(u|x, u') \tilde D_{h_t}(u|x,t) d u
\to  \int_{\mathcal{U}} g(u|x, u') \tilde D_{h}(u|x)    du
\end{equation}
uniformly in $(u',x) \in \mathcal{U}  \mathcal{X}$,
for any measurable $g$ such that $\sup_{u',x}
|g(u|x,u')| \in \ell^{1}(\mathcal{U})$ and such that
$(x,u')\mapsto g(u|x, u')$ is  continuous  for a.e.
$u$.
\end{proposition}

It is important to note that Proposition \ref{proposition:
hadamard_diff linear_funct} applies to integrals defined over entire
domains, unlike Proposition \ref{proposition: hadamard_diff} which
states uniform convergence of integrands over domains excluding
non-regular neighborhoods. (Thus,  Proposition 3 does not
immediately follow from Proposition 2.)    Here integration acts
like a smoothing operation and allows us to ignore these non-regular
neighborhoods. In order to prove convergence of integrals defined
over entire domains, we couple the almost everywhere convergence implied by
Proposition \ref{proposition: hadamard_diff} with the uniform
integrability of Lemma 3 in the Appendix, and then interchange
limits and integrals. We should also note that an alternative way of
proving result (\ref{equation: had_diff_lin_fun_part2}), but
\textit{not}
 other results in the paper, can be based on the convexity of
the functional in (\ref{equation: had_diff_lin_fun_part2}) with
respect to the underlying curve, following the approach of Mossino
and Temam (1981), and Alvino et al. (1989). Due to this limitation,
we do not pursue this approach in this paper.

It is also worth emphasizing the properties of the following
smoothed functionals.  For a measurable function $f: \Bbb{R} \mapsto
\Bbb{R},$ define the smoothing operator $S$ as
\begin{equation}\label{smooth}
S f(y') : = \int k_{\delta}(y'-y) f(y) dy,
 \end{equation}
where $k_{\delta}(v) = 1\{ |v| \leq \delta \}/2\delta$ and
$\delta>0$ is a fixed bandwidth. Accordingly, the smoothed curves
$SF$ and $ SQ^{\ast}$ are given by
\begin{equation*}
SF(y'|x) :=  \int k_{\delta}(y'-y) F(y|x) d y,  \ \ \ \
SQ^{\ast}(u'|x) := \int k_{\delta}(u'-u) Q^{\ast}(u|x) d u.
\end{equation*}
Note that given the quantile function $Q^*$, the smoothed function
$SQ^*$ has a convenient interpretation as a local average quantile
function or fractile. Since we form these curves as differences of
the elementary functionals in Proposition \ref{proposition:
hadamard_diff linear_funct} divided by $2\delta$, the following
corollary is immediate:

\begin{corollary}[Hadamard derivative of smoothed $Q^\ast$ and $F$
with respect to $Q$] We have that $ SD_{h_t}(y'|x,t)  \to
SD_{h}(y'|x) $ uniformly in $(y',x) \in \mathcal{YX}$, and $ S\tilde
D_{h_t}(u'|x,t) \to S\tilde D_{h}(u'|x) $ uniformly in $(u',x) \in
\mathcal{UX}$.  The results hold uniformly in the smoothing
parameter $\delta \in [\delta_1, \delta_2]$, where $\delta_1$ and
$\delta_2$ are positive constants.
\end{corollary}
Note that smoothing allows us to achieve uniform convergence over the
entire domain,  without excluding non-regular neighborhoods.

\subsection{Empirical Properties and Functional Limit Theory for Rearranged Estimators}
Here we state a finite sample result and then derive functional
limit laws for rearranged estimators.  These results constitute the
second set of original main theoretical results obtained in this
paper.

The following  proposition shows that the rearranged quantile curves have
smaller estimation error than the original curves whenever the latter are not
monotone.

\begin{proposition}[Improvement in estimation property provided by
 rearrangement]
Suppose that $\widehat{Q}$ is an estimator (not necessarily
consistent) for some true quantile curve $Q_0$. Then, the rearranged
curve $ \widehat{Q}^{\ast}$ is closer to the true curve than
$\widehat{Q}$ in the sense that, for each $x \in \mathcal{X}$,
\begin{equation*}
\| \widehat Q^{\ast} - Q_0 \|_{p}  \leq \| \widehat Q - Q_0 \|_{p},
\ p \in [1, \infty],
\end{equation*}
where $\| \cdot \|_p$ denotes the $L^p$ norm of a measurable
function $Q: \mathcal{U} \mapsto \mathbb{R}$, namely $\| Q \|_p = \{
\int_\mathcal{U} |Q(u)|^p du \}^{1/p}$. The inequality is strict for
$p\in(1,\infty)$ whenever $u \mapsto \widehat{Q}(\protect u |x)$ is
strictly decreasing on a subset of $\mathcal{U}$ of positive
Lebesgue measure, while $u \mapsto Q_0(u |x)$ is strictly increasing
on $\mathcal{U}$. The above property is independent of the sample
size and of the way the estimate of the curve is obtained, and thus
continues to hold in the population.
\end{proposition}

This property suggests that the rearranged estimators should be
preferred over the original estimators.   Moreover, this property
does not depend on the way the quantile model is estimated or any
other specifics, and is thus applicable quite generally.  Regarding
the proof of this property, the weak reduction in estimation error
follows from an application of a classical rearrangement inequality
of Lorentz (1953) and the strict reduction follows from its
appropriate strengthening (Chernozhukov et al.,
2009).\footnote{Similar contractivity properties have been shown for
the pool adjacent violators algorithm in different contexts. See,
for example, Robertson et al. (1988) for isotonic regression, and
Eggermont and LaRiccia (2000) for monotone density estimation. Glad
et al. (2003) shows that a density estimator corrected to be a
proper density satisfies a similar property.}

In order to derive the functional limit laws of rearranged estimators, we maintain the following assumptions on
$\widehat Q$ throughout the paper:

\noindent \textbf{Assumption 2.} (Properties of $\widehat Q$).
\textit{The empirical curve $\widehat{Q}$ takes its values in the
space of bounded measurable functions defined on
$\mathcal{U}\mathcal{X}$, and, in $\ell ^{\infty
}(\mathcal{U}\mathcal{X})$,
\begin{equation}\label{eq: FCLT}
a_n (\widehat{Q}(u|x )- Q(u|x ))\Rightarrow G( u |x),
\end{equation}
as a stochastic process indexed by $(u,x) \in
\mathcal{U}\mathcal{X}$, where $ (u,x) \mapsto G(u|x )$ is a
stochastic process (typically Gaussian) with continuous paths. Here
$a_n$ is a sequence of constants such that $a_n \to \infty$ as $n
\to \infty$, where $n$ is the sample size.}

This assumption requires that the original quantile estimator
satisfies a functional central limit theorem with a continuous limit
stochastic process over the domain $\mathcal{U}=[0,1]$  for the index
$u$. If (\ref{eq: FCLT}) holds only over a subinterval of $[0,1]$,
we can accommodate the reduced domain following Remark 1. This key
condition is rather weak and holds for a wide variety of conditional
and structural quantile estimators.\footnote{For sufficient
conditions, see, for example, Gutenbrunner and Jure\v{c}kov\'{a}
(1992), Portnoy (1991), Angrist et al. (2006), and Chernozhukov and
Hansen (2006).} With an appropriate normalization rate and a fixed
$x$,   this assumption holds for series quantile regressions. For
example, Belloni and Chernozhukov (2007) extended the results of He
and Shao (2000) to the process case and established the functional
central limit theorem for $a_n(\widehat Q(u|x) - Q(u|x))$ for a
fixed $x$. At the same time, we should also point out that this
assumption does not need to hold in all estimation problems with
monotonicity restrictions.\footnote{ For example, Assumption 2 does
not hold when we estimate monotone production or demand functions $u
\mapsto f(u)$, where $u$ is input or price, using nonparametric
kernel or series regression. We refer the reader to Chernozhukov et.
al. (2009) for appropriate further results that enable us to perform
uniform inference in such cases. On the other hand, Assumption 1
does hold when we estimate monotone production or demand functions
$u \mapsto f(u)$ using parametric or semi-parametric regression. }

The following proposition derives functional limit laws for the
rearranged quantile estimator $\widehat Q^\ast$ and the
corresponding distribution estimator $\widehat F$, using the
functional differentiation results for the rearrangement-related operators from
the previous section.

\begin{proposition}[Functional limit laws for $\widehat{F}$ and $\widehat{Q}^{\ast}$]\label{proposition: limit_distribution}
In $\ell ^{\infty }(K)$, where $K$ is any compact subset of
$\mathcal{YX}^{\ast }$,
\begin{equation}\label{eq: FCLT1}
a_n(\widehat{F}(y|x)-F(y|x))\Rightarrow D_{G}(y|x)
\end{equation}%
as a stochastic process indexed by $(y,x)\in \mathcal{YX}^{\ast }$;
and in $\ell ^{\infty }(\mathcal{UX}_{K})$, with $\mathcal{UX}_{K} =
\{( u,x) : (Q^{\ast}( u |x),x)\in K \}$,
\begin{equation}\label{eq: FCLT2}
a_n(\widehat{Q}^{\ast}( u |x)-Q^{\ast}( u |x))\Rightarrow \widetilde
D_{G}(u|x),
\end{equation}
as a stochastic process indexed by $(u,x)\in \mathcal{UX}_{K}$; where  the maps
$h \mapsto D_h$ and $h \mapsto \widetilde D_h$ are defined in equations (\ref{D1b}) and (\ref{D2b}).
\end{proposition}
This proposition provides the basis for inference using rearranged
quantile estimators and corresponding distribution estimators.

Let us first discuss  inference for the case with a monotonic population
curve $Q$.  It is useful to emphasize the following corollary of Proposition 5:
\begin{corollary}[Functional limit laws for $\widehat{F}$ and $\widehat{Q}^{\ast}$ in the monotonic case]\label{cor: monotone}  Suppose $u\mapsto Q(u|x)$ has $\partial_u Q(u|x)> 0$ for each
$(u,x)\in \mathcal{UX}$. Then  $\mathcal{YX}^* = \mathcal{YX}$ and
$\mathcal{UX}^* = \mathcal{UX}$. Accordingly, the convergence in
Proposition \ref{proposition: limit_distribution} holds uniformly
over  the entire $\mathcal{YX}$ and $\mathcal{UX}$. Moreover,
$\widetilde D_{G}(u|x) = G(u|x)$, i.e., the rearranged quantile
curves have the same first order asymptotic distribution as the
original estimated quantile curves.
\end{corollary}
Thus, if the population curve is monotone, we can rearrange the
original non-monotone quantile estimator to be monotonic
\textit{without} affecting its (first order) asymptotic properties.
Hence, all the inference tools that apply to the original quantile
estimator $\widehat Q$ also apply to the rearranged quantile
estimator $\widehat Q^\ast$. In particular, if the bootstrap is
valid for the original estimator, it is also valid for the
rearranged estimator, by the functional delta method for the
bootstrap.  Thus, when $Q$ is monotone, Corollary \ref{cor:
monotone}, enables us to perform uniform inference on $Q$ and $F$
based on the rearranged estimators $\widehat Q^\ast$ and $\widehat
F$.

\textbf{Remark 2.}(\textit{Detecting and avoiding cases with non-monotone
$Q$.}) Before discussing inference for the case with a non-monotonic
population curve $Q$, let us first emphasize that since
non-monotonicity of $Q$ is a rather obvious sign of specification
error, it is best to try to detect and  avoid this case. For this
purpose we should use sufficiently flexible functional forms and
reject the ones that fail to pass monotonicity tests. For example,
we can use the following generic test of monotonicity for $Q$: If
$Q$ is monotone, the first order behavior of $\widehat Q^*$ and
$\widehat Q$ coincides, and if $Q$ is not monotone, $\widehat Q^*$
and $\widehat Q$ converge to different probability limits $Q^*$ and
$Q$. Therefore, we can reject the hypothesis of monotone $Q$ if a
uniform confidence region for $Q$ based on $\widehat Q$ does not
contain $\widehat Q^*$, for at least one point $x \in
\mathcal{X}$.\footnote{This test is conservative, but it is generic
and very inexpensive.   In order to build non-conservative tests, we
need to derive the limit laws for $\|\widehat Q - \widehat Q^*\|$
for suitable norms $\|\cdot\|$. These laws will depend on
higher-order functional limit laws for quantile estimators, which
appear to be non-generic and have to be dealt with on a case by case
basis. }  \qed

Let us now discuss  inference for the case with a non-monotonic
population curve $Q$. In this case, the large sample properties of
the rearranged quantile estimators $\widehat Q^*$ substantially
differ from those of the initial quantile estimators $\widehat Q$.
Proposition \ref{proposition: limit_distribution} still enables us
to perform uniform inference on the rearranged population curve
$Q^*$ based on the rearranged estimator $\widehat Q^*$, but only
after excluding certain nonregular neighborhoods (for the
distribution estimators, the neighborhoods of the critical values of
the map $u \mapsto Q(u|x)$, and, for the rearranged quantile
estimators, the image of these neighborhoods under $F$). These
neighborhoods can be excluded by locating the points $(u,x)$ where a
consistent estimate of $|\partial_u Q(u|x)|$  is close to zero; see
Hendricks and Koenker (1991) for a consistent estimator of
$|\partial_u Q(u|x)|$.

Next we consider the following linear functionals of the rearranged quantile
and distribution estimates:
\begin{equation*}\begin{split}
&  (y',x)\mapsto \int_{\mathcal{Y}} g(y|x, y') \widehat F(y|x) dy, \
\ (u',x)\mapsto \int_{\mathcal{U}} g(u|x, u')
\widehat{Q}^{\ast}(u|x) d u.
\end{split}\end{equation*}
The following proposition derives functional limit laws for  these
functionals.\footnote{ Working with these functionals is equivalent
to placing our empirical processes into the space $L^p$ ($p=1$ for
rearranged distributions and $p=\infty$ for quantiles), equipped
with weak* topology, instead of strong topology. Convergence in law
of the integral functionals, shown in Proposition \ref{proposition:
limit_distribution_linear_func}, is equivalent to the convergence in
law of the rearranged estimated processes in such a metric space.}
Here the convergence results hold without excluding any nonregular
neighborhoods, which is convenient for practice in the non-monotonic
case.

\begin{proposition}[Functional limit laws for linear functionals of $\widehat Q^\ast$ and $\widehat F$]
\label{proposition: limit_distribution_linear_func} Under
the same restrictions on the function $g$ as in Proposition
\ref{proposition: hadamard_diff linear_funct}, the following results
hold with the limits being continuous on the specified domains:
\begin{equation}\label{eq: FCLT3}
1. \quad a_n \int_{\mathcal{Y}} g(y|x, y') (\widehat{F}(y|x)-F(y|x))
d y \Rightarrow \int_{\mathcal{Y}}  g(y|x, y') D_{G}(y|x) dy,
\end{equation}
as a stochastic process indexed by $(y',x)\in \mathcal{Y}
\mathcal{X}$,  in $\ell^{\infty}(\mathcal{Y}   \mathcal{X})$.
 \begin{equation}\label{eq: FCLT4}
2. \quad a_n  \int_{\mathcal{U}} g(u|x, u') (\widehat{Q}^{\ast}( u
|x)-Q^{\ast}( u |x)) d u \Rightarrow \int_{\mathcal{U}}  g(u|x, u')
\tilde D_{G}(u|x) du,
 \end{equation}
as a stochastic process indexed by $(u',x) \in \mathcal{U}
\mathcal{X}$,  in $\ell^{\infty}(\mathcal{U}  \mathcal{X})$.

\end{proposition}

The linear functionals defined above
 are useful building blocks for various statistics,
such as partial means, various moments, and Lorenz curves. For
example, the conditional Lorenz curve based on rearranged quantile
estimators is
 \begin{equation}\label{eq: Lorenz}
\widehat L(u'|x) : = \Big ( \int_{\mathcal{U}}1 \{ u \leq u' \}
\widehat{Q}^{\ast}(u|x) d u \Big) /  \Big ( \int_{\mathcal{U}}
\widehat{Q}^{\ast}(u|x)  d u \Big),
 \end{equation}
which is a ratio of partial and overall conditional means. Hadamard
differentiability of the mapping
 \begin{equation}\label{eq: ratio}
Q \mapsto  L(u'|x) : = \Big ( \int_{\mathcal{U}}1 \{ u \leq u' \}
{Q}^{\ast}(u|x) d u \Big) /  \Big ( \int_{\mathcal{U}}
{Q}^{\ast}(u|x)  d u \Big),
 \end{equation}
with respect to $Q$ immediately follows from (a) the
differentiability of a ratio $ \beta/\gamma$ with respect to its
numerator $\beta$ and denominator $\gamma$ at $\gamma \neq 0$, (b)
Hadamard differentiability of the numerator and denominator in
(\ref{eq: ratio}) with respect to $Q$ established in Proposition
\ref{proposition: limit_distribution_linear_func}, and (c) the chain
rule for the Hadamard derivative.  Hence, provided that $Q>0$ so
that $Q^*>0$, we have that in the metric space
$\ell^{\infty}(\mathcal{UX})$
 \begin{equation}\label{eq: FCLT Lorenz}
 a_n(\widehat L(u'|x)- L(u'|x)) \Rightarrow L(u'|x)\cdot \left ( \frac{  \int_{\mathcal{U}}  1 \{ u \leq u' \}
\tilde D_{G}(u|x) du }{ \int_{\mathcal{U}}1 \{ u \leq u' \}
{Q}^{\ast}(u|x) d u   } - \frac {    \int_{\mathcal{U}} \tilde
D_{G}(u|x) du   }{ \int_{\mathcal{U}} {Q}^{\ast}(u|x)  d u } \right
),
 \end{equation}
as an empirical process indexed by $(u',x) \in \mathcal{UX}$. In
particular, validity of the bootstrap for estimating this functional
limit law in (\ref{eq: FCLT Lorenz}) holds by the functional delta
method for the bootstrap.

We next consider the empirical properties of the smoothed curves
obtained by applying the linear smoothing operator $S$ defined in
(\ref{smooth}) to $\widehat F$ and $\widehat{Q}^{\ast}$:
\begin{equation*}
S\widehat F(y'|x) :=  \int k_{\delta}(y'-y) \widehat F(y|x) d y, \ \
\ \ S\widehat{Q}^{\ast}(u'|x) := \int k_{\delta}(u'-u)
\widehat{Q}^{\ast}(u|x) d u.
\end{equation*}
The following corollary immediately follows from
Corollary 2 and the functional delta method.

\begin{corollary}[Functional limit laws for smoothed $\widehat Q^\ast$ and $\widehat F$]
In $\ell ^{\infty}(\mathcal{YX})$,
\begin{equation}\label{eq: FCLT5}
a_n(S\widehat{F}(y'|x)-SF(y'|x))\Rightarrow SD_{G}(y'|x),
\end{equation}%
as a stochastic process indexed by $(y',x) \in \mathcal{YX}$, and in
$\ell ^{\infty }(\mathcal{UX})$,
\begin{equation}\label{eq: FCLT6}
a_n(S\widehat{Q}^{\ast}( u' |x)-SQ^{\ast}( u' |x))\Rightarrow
S\widetilde D_{G}(u'|x),
\end{equation}
as a stochastic process indexed by $(u',x) \in \mathcal{UX}$. The
results hold uniformly in the smoothing parameter $\delta \in
[\delta_1, \delta_2]$, where $\delta_1$ and $\delta_2$ are positive
constants.
\end{corollary}

Thus, as in the case of linear functionals, we can perform inference
on $SQ^{\ast}$ based on the smoothed rearranged estimates without
excluding nonregular neighborhoods, which is convenient for practice
in the non-monotonic case. Furthermore, validity of the bootstrap
for the smoothed curves follows by the functional delta method for
the bootstrap. Lastly, we note that it is not possible to
simultaneously allow $\delta \to 0$ and preserve the uniform
convergence stated in the corollary.

Our final corollary asserts validity of the bootstrap for inference
on rearranged estimators and their functionals.  This corollary
follows from the functional delta method for the bootstrap (e.g.,
Theorem 13.9 in van der Vaart, 1998).

\begin{corollary}[Validity of the bootstrap for estimating laws of rearranged estimators] If the bootstrap consistently estimates the functional limit law (\ref{eq: FCLT}) of the empirical process
$\{a_n (\widehat Q(u|x) - Q(u|x), (u,x) \in \mathcal{UX}\}$, then it also
consistently estimates the functional limit laws (\ref{eq: FCLT1}), (\ref{eq:
FCLT2}), (\ref{eq: FCLT3}), (\ref{eq: FCLT4}), (\ref{eq: FCLT Lorenz}), (\ref{eq: FCLT5}), and
(\ref{eq: FCLT6}).
\end{corollary}

\section{Examples}

In this section we apply rearrangement to the estimation of
structural quantile and distribution functions.  We show how
rearrangement monotonizes instrumental quantile and distribution
function estimates, and demonstrate how to perform inference on the
target functions using the results developed in this paper. Using a
supporting numerical example, we show that rearranged estimators
noticeably improve upon original estimators and also outperform
isotonized estimators.  Thus, rearrangement is necessarily
preferable to the standard approach of  simply ignoring
non-monotonicity. Moreover, in quantile estimation problems,
rearrangement is also preferable to the standard approach of
isotonization used primarily in mean estimation problems.

\subsection{Empirical Example} 
\begin{figure}

\begin{center}

\centering\epsfig{figure=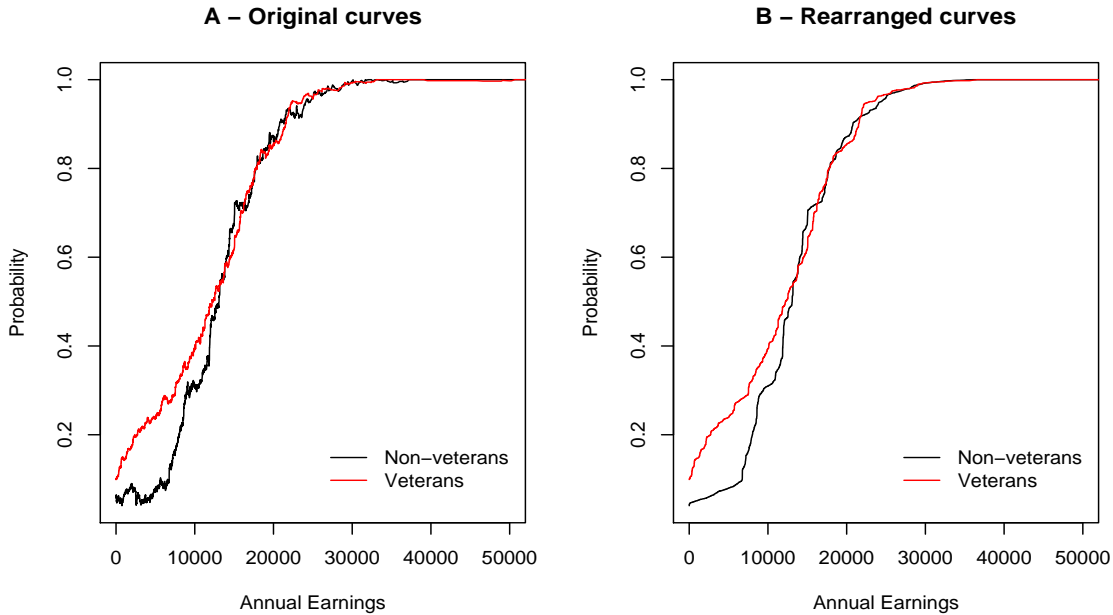,width=6in,height=3.5in}

\caption{\label{Fig:5} Abadie's estimates  of the structural
distributions of earnings  for veteran and non-veterans (left
panel), and their rearrangements (right panel).}

\end{center}

\end{figure}

\begin{figure}

\begin{center}

\centering\epsfig{figure=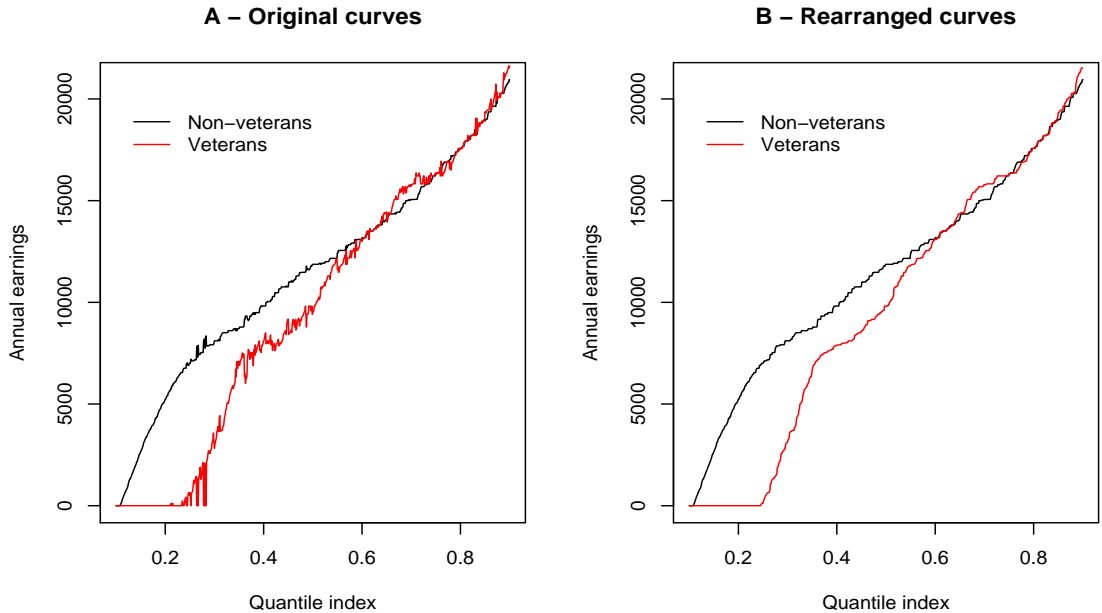,width=6in,height=3.5in}
\caption{\label{Fig:6} Chernozhukov and Hansen's estimates of the
structural quantile functions of earnings for veterans (left panel),
and their rearrangements (right panel). }

\end{center}

\end{figure}

\begin{figure}
\begin{center}

\centering\epsfig{figure=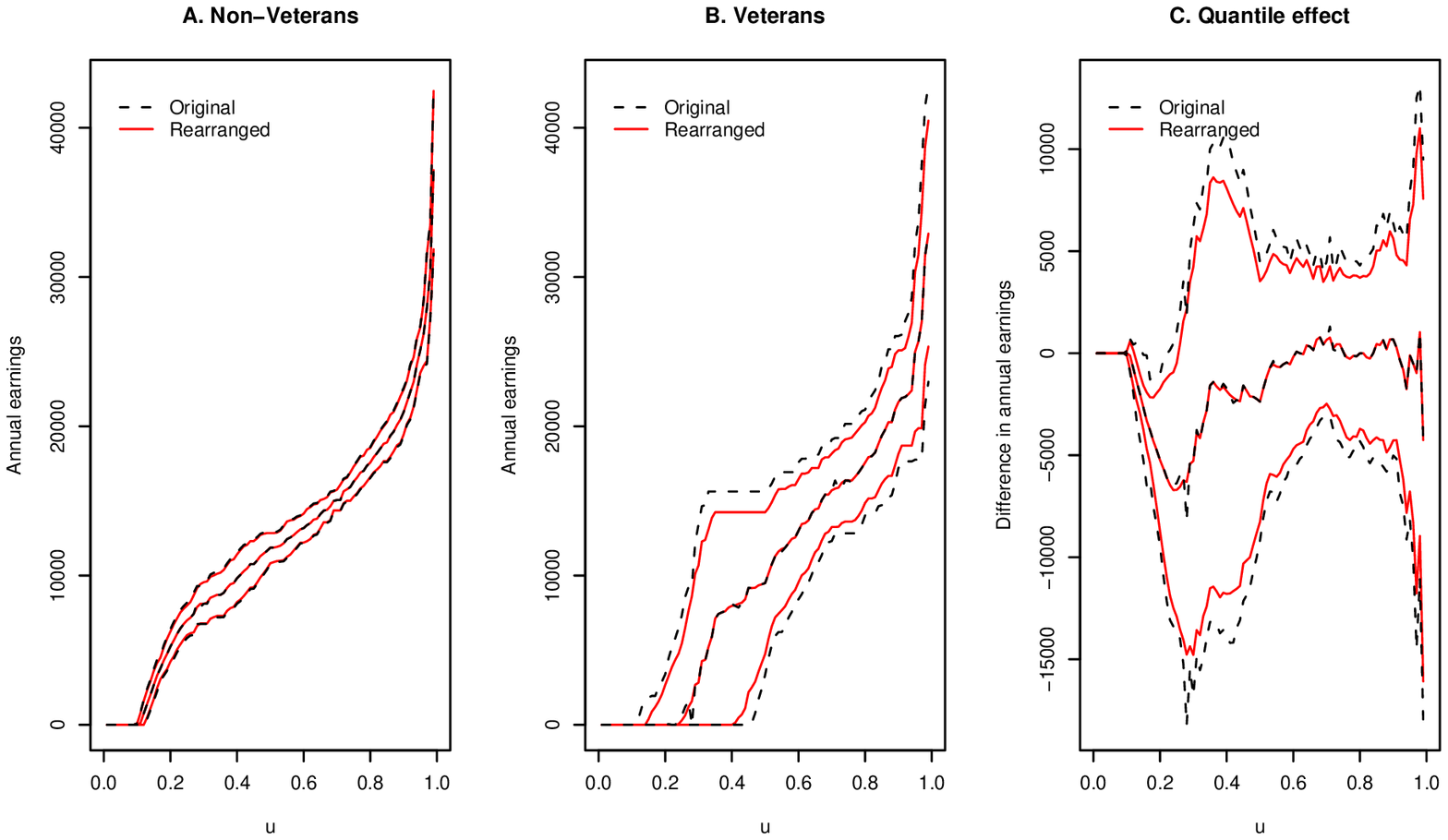,width=5in,height=6in, angle=-90}

\caption{\label{Fig:7}  Original and rearranged point estimates and corresponding simultaneous 90\% confidence bands for
structural quantile functions of earnings (panels A  and B) and structural quantile
effect of Vietnam veteran status on earnings (panel C).  The bands for the
quantile functions (panels A  and B) are intersected with the class of monotone
functions.}

\end{center}
\end{figure}

We consider estimation of the causal/structural effects of Vietnam veteran
status $X\in \{0,1\}$ in the quantiles and distribution of civilian
earnings $Y$. Since veteran status is likely to be endogenous
relative to potential civilian earnings, we employ an instrumental
variables approach, using the U.S. draft lottery as an instrument
for the Vietnam status (Angrist, 1990). We use the same data subset
from the Current Population Survey as in Abadie
(2002).\footnote{These data consist of a sample of white men, born
in 1950--1953, from the March Current Population Surveys of 1979 and
1981-1985. The data include annual labor earnings, the Vietnam
veteran status and an indicator on the Vietnam era lottery. There
are 11,637 men in the sample, with 2,461 Vietnam veterans and 3,234
eligible for U.S. military service according to the draft lottery
indicator. Abadie (2002) gives additional information on the data
and the construction of the variables.} We then estimate structural
quantile and distribution functions with the instrumental quantile
regression estimator of Chernozhukov and Hansen (2005, 2006) and the
instrumental distribution regression estimator of Abadie (2002).
Under some assumptions these procedures consistently estimate the
structural quantile and distribution functions of
interest.\footnote{More specifically, Abadie's (2002) procedure
consistently estimates these functions for the subpopulation of
compliers under instrument independence and monotonicity.
Chernozhukov and Hansen's  (2005, 2006) approach consistently
estimates these functions for the entire population under instrument
independence and rank similarity.} However, like most estimation
methods mentioned in the introduction, neither of these procedures
explicitly imposes monotonicity of the distribution and quantile
functions. Accordingly, they can produce estimates in finite samples
that are nonmonotonic due to either sampling variation or violations
of instrument independence or other modeling assumptions. We
monotonize these estimates using rearrangement and perform inference
on the target structural functions using uniform confidence bands
constructed via bootstrap. We use the programming language R to
implement the procedures (R Development Core Team, 2007). We present
our estimation and inference results in Figures
\ref{Fig:5}--\ref{Fig:7}.

In Figure \ref{Fig:5}, we show Abadie's estimates of the structural
distribution of earnings for veterans and non-veterans  (left panel)
as well as their rearrangements (right panel).   For both veterans
and non-veterans, the original estimates of the distributions
exhibit clear local non-monotonicity. The rearrangement fixes this
problem producing increasing estimated distribution functions. In
Figure \ref{Fig:6}, we show Chernozhukov and Hansen's estimates of
the structural quantile functions of earnings for veterans (left
panel) as well as their rearrangements (right panel). For both
veterans and non-veterans, the estimates of the quantile functions
exhibit pronounced local non-monotonicity. The rearrangement fixes
this problem  producing increasing estimated quantile functions . In
the case of quantile functions, the nonmonotonicity problem is
specially acute for the small sample of veterans.

In Figure \ref{Fig:7}, we plot uniform $90\%$ confidence bands for
the structural quantile functions of earnings for veterans and
non-veterans, together with uniform $90\%$ confidence bands for the
effect of Vietnam veteran status on the quantile functions for
earnings, which measures the difference between the structural
quantile functions for veterans and non-veterans. We construct the
uniform confidence bands using both the original estimators and the
rearranged estimators based on 500 bootstrap repetitions and a fine
net of quantile indices $\{0.01, 0.02, ...,0.99 \}$. We obtain the
bands for the rearranged functions assuming that the population
structural quantile regression functions are monotonic, so that the
first order behavior of the rearranged estimators coincides with the
behavior of the original estimators. The figure shows that even for
the large sample of non-veterans the rearranged estimates lie within
the original bands, thus passing our automatic test of monotonicity
specified in Remark 2. Thus, the lack of monotonicity of the
estimated quantile functions in this case is likely caused by
sampling error. From the figure, we conclude that veteran status has
a statistically significant negative effect in the lower tail, with
the bands for the rearranged estimates showing a wider range of
quantile indices for which this holds.


\subsection{Monte Carlo} We design a Monte Carlo experiment to closely match the previous empirical example.  In particular, we consider
a location model, where the outcome is $Y= [1, X] \alpha +
\epsilon$, the endogenous regressor is $X = 1 \{ [1; Z] \pi + v \geq
0 \},$ the instrument $Z$ is a binary random variable, and the
disturbances $(\epsilon, v)$ are jointly normal and independent of
$Z$. The true structural quantile functions are $ Q_0(u|x) = [1;x]
\alpha + Q_\epsilon(u),  \ x \in \{0,1\},$ where $Q_{\epsilon}$ is
the quantile function of the normal variable $\epsilon$. The
corresponding structural distribution functions are the inverse of
the quantile functions with respect to $u$.  We select the value of
the parameters by estimating this location model parametrically by
maximum likelihood, and  then generate samples from the estimated
model, holding the values of the instruments $Z$ equal to those in
the data set.\footnote{More specifically, after normalizing the
standard deviation of $v$ to one, we set $\pi = [-.92;
.40]^\mathsf{T}$, $\alpha = [11,753; -911]^\mathsf{T}$, the standard
deviation of $\epsilon$ to $8,100$, and the covariance between
$\epsilon$ and $v$ to $379$.  We draw $5,000$ Monte Carlo samples of
size $n=11,627$. We generate the values of $Y$ and $X$ by drawing
disturbances  $(\epsilon, v)$ from a bivariate normal distribution
with zero mean and the estimated covariance matrix.} We use the
estimators for the structural distribution and quantile functions
described in the previous section.  We monotonize the estimates
using either rearrangement or isotonization.  We use isotonization
as a benchmark since it is the standard approach in mean regression
problems (Mammen, 1991); it amounts to projecting the estimated
function on the set of monotone functions.

Table 1 reports ratios of estimation errors of the rearranged and
isotonized estimates to those of the original estimates, recorded in
percentage terms. The target functions  are the structural
distribution and quantile functions. We measure estimation errors
using the average $L^p$ norms $\| \cdot \|_p$ with $p=1,2,$ and
$\infty$, and we compute them as Monte Carlo averages of $\|f_0 -
\tilde f \|_p,$  where  $f_0$ is the target function, and $\tilde f$
is either the original or rearranged or isotonized estimate of this
function.

We find that the rearranged estimators noticeably outperform the
original estimators, achieving a reduction in estimation error up to
$14\%$, depending on the target function and the norm. Moreover, in
this case the better approximation of the rearranged estimates to
the structural functions also produces more accurate estimates of
the distribution and quantile effects, achieving a $3\%$ to $9\%$
reduction in estimation error for the distribution estimator and a
$3\%$ to $14\%$ reduction in estimation error for the quantile
estimator, depending on the norm.

We also find that the rearranged estimators  outperform the
isotonized estimators, achieving up to a further $4\%$ reduction in
estimation error, depending on the target function and the norm. The
reason is that isotonization projects the original fitted function
on the set of monotone functions, converting non-monotone segments
into flat segments. In contrast, rearrangement sorts the original
fitted function,  converting non-monotone segments into steep,
increasing segments that preserve measure.  In the context of
estimating quantile and distribution functions, the target functions
tend to be non-flat, suggesting that rearrangement should be
typically preferred over isotonization.\footnote{To give some
intuition about this point, it is instructive to consider a simple
example with a two-point domain $\{0,1\}$. Suppose that the target
function $f_0: \{0, 1\} \to \Bbb{R}$ is increasing, and steep,
namely $f_0(0) > f_0(1)$, and the fitted function $\widehat f : \{0,
1\} \to \Bbb{R}$ is decreasing, with $\widehat f(0) > \widehat
f(1)$. In this case, isotonization produces a nondecreasing function
$\bar f : \{0, 1\} \to \Bbb{R}$, which is flat, with $\bar f(0) =
\bar f(1) = [\widehat f(0) + \widehat f (1)]/2 $, which is somewhat
unsatisfactory. In such cases rearrangement can significantly
outperform isotonization, since it produces the steepest fit, namely
it produces   $\widehat f^* : \{0, 1\} \to \Bbb{R} $ with $ \widehat
f^*(0) = \widehat f(1) < \widehat f^*(1) = \widehat f(0)$. This
observation provides a simple theoretical underpinning for the
estimation results we see in Table 1.}

\begin{table}[hptb]
\vspace{3mm} \caption{Ratios of estimation error of rearranged and
isotonic estimators to those of original estimators, in percentage terms.}

\begin{center}
\begin{tabular}{lcccccccc} \hline\hline
\multicolumn{1}{l}{ }&\multicolumn{2}{c}{ Veterans }&\multicolumn{1}{c}{  }&\multicolumn{2}{c}{ Non-Veterans}&\multicolumn{1}{c}{  }&\multicolumn{2}{c}{ Effect} \\
\cline{2-3} \cline{5-6} \cline{8-9} \multicolumn{1}{l}{
}&\multicolumn{1}{c}{Rearranged}&\multicolumn{1}{c}{Isotonized}&\multicolumn{1}{c}{}&\multicolumn
{1}{c}{Rearranged} &\multicolumn {1}{c}{Isotonized}
&\multicolumn{1}{c}{}&\multicolumn{1}{c}{Rearranged}&\multicolumn{1}{c}{Isotonized}
\\ \hline
\multicolumn{1}{l}{}&\multicolumn{8}{c}{Structural Distribution function}
\\
$L^1$       &99  &99  &  &97 &98 &  &97  &98  \\
$L^2$       &99  &99  &  &97 &98 &  &97  &99  \\
$L^{\infty}$&96  &98  &  &90 &94 &  &91  &95  \\
\multicolumn{1}{l}{}&\multicolumn{8}{c}{Structural Quantile function}
\\
$L^1$       &97  &98  &  &100 &100 &  &97  &98  \\
$L^2$       &96  &97  &  &100 &100 &  &96  &98  \\
$L^{\infty}$&86  &87  &  &98  &99  &  &86  &88
\\
\hline\hline
\end{tabular}
\end{center}
\end{table}

\section{Conclusion}

This paper develops a  monotonization procedure for estimation of
conditional and structural quantile and distribution functions based
on rearrangement-related operations. Starting from a possibly
non-monotone empirical curve, the procedure produces a rearranged
curve that not only satisfies the natural monotonicity requirement,
but also has smaller estimation error than the original curve. We
derive asymptotic distribution theory for the rearranged curves, and
illustrate the usefulness of the approach with an empirical
application and a simulation example. There are many potential
applications of the results given in this paper and companion work
(Chernozhukov et al., 2009) to other econometric problems with shape
restrictions (see e.g. Matzkin, 1994).

\appendix

\section{Proofs}

\subsection{Proof of Proposition 1} First, note that the distribution of $%
Y_{x}$ has no atoms, i.e.,
\begin{equation*}
\Pr[Y_{x}=y]=\Pr[Q(U|x)=y]=\Pr[U\in \{u \in \mathcal{U}: u \text{ is
a root of } Q( u |x)=y\}]=0,
\end{equation*}%
since the number of roots of $Q( u |x)=y$ is finite under Assumption
1, and $U \sim \text{Uniform}(\mathcal{U})$.
Next, by Assumption 1 the number of critical values of $%
Q(u|x )$ is finite, hence claim (1) follows.

Next, for any regular $y$, we can write $F(y|x)$ as
\begin{align}\begin{split}\nonumber
\int_0^1 1\{Q(u|x) \leq y\} du & =    \sum_{k=0}^{K(y|x)-1}
\int_{u_k(y|x)}^{u_{k+1}(y|x)} 1\{Q(u|x) \leq y\}d u  +
\int_{u_{K(y|x)}(y|x)}^1 1\{Q(u|x) \leq y\}d u,
\end{split}\end{align}
where $ u_{0}(y|x):=0$ and $\{ u _{k}(y|x),\text{ for }k=1,2,
...,K(y|x)<\infty \}$ are the roots of $Q( u |x)=y$ in increasing
order. Note that the sign of $\partial_u Q(u|x)$ alternates over
consecutive $ u _{k}(y|x)$, determining whether $1\{Q(y|x) \leq
y\}=1$ on the interval $[u_{k-1}(y|x), u_k(y|x)]$. Hence the first
term in the previous expression simplifies to $
\sum_{k=0}^{K(y|x)-1} 1\{\partial_u Q(u_{k+1}(y|x)|x) \geq 0
\}(u_{k+1}(y|x) - u_{k}(y|x)); $ while the last term simplifies to $
1\{\partial_u Q(u_{K(y|x)}(y|x)|x) \leq 0 \} ( 1 - u_{K(y|x)}(y|x)).
$ An additional simplification yields the expression given in claim
(2) of the proposition.

The proof of claim (3) follows by taking the derivative of
expression in claim (2), noting that at any regular value $y$ the
number of solutions $K(y|x)$ and $\textrm{sign}(\partial_u Q( u
_{k}(y|x)|x))$ are locally constant; moreover,
\begin{equation*}
 \partial_y u _{k}(y|x)= \frac{\textrm{sign}(\partial_u Q( u _{k}(y|x)|x))}{|\partial_u Q( u _{k}(y|x)|x)|}.
\end{equation*}
Combining these facts we get the expression for the derivative given
in claim (3).

To show the absolute continuity of $F$ with $f$ being the
Radon-Nykodym derivative, it suffices to show that for each $y' \in
\mathcal{Y}_{x}$, $\int^{y'}_{-\infty} f(y|x) dy =
\int_{-\infty}^{y'} d F(y|x)$, cf. Theorem 31.8 in Billingsley
(1995). Let $V_{t}^{x}$ be the union of closed balls of radius $t$
centered on the critical points $\mathcal{Y}_{x}\setminus
\mathcal{Y}_{x}^{\ast }$, and define
$\mathcal{Y}_{x}^{t}=\mathcal{Y}_{x}\backslash V_{t}^{x}$. Then,
$\int_{-\infty}^{y'} 1\{y \in \mathcal{Y}_{x}^t\} f(y|x) dy =
\int_{-\infty}^{y'} 1\{y \in \mathcal{Y}_{x}^t\} d F(y|x)$. Since
the set of critical points $\mathcal{Y}_{x}\setminus
\mathcal{Y}_{x}^{\ast }$ is finite and has mass zero under $F$,
$\int_{-\infty}^{y'} 1\{y \in \mathcal{Y}_{x}^t\} d F(y|x) \uparrow
\int_{-\infty}^{y'} d F(y|x)$ as $t \to 0$. Therefore,
$\int_{-\infty}^{y'} 1\{y \in \mathcal{Y}_{x}^t\} f(y|x) dy \uparrow
\int_{-\infty}^{y'} f(y|x) dy= \int_{-\infty}^{y'}d F(y|x).$

Claim (4) follows by noting that at the regions where $s \mapsto
Q(s|x)$ is increasing and one-to-one,  we have that  $
F(y|x)=\int_{Q(s|x)\leq y}ds=\int_{s\leq
Q^{-1}(y|x)}ds=Q^{-1}(y|x)$. Inverting the equation $ u =F(Q^{\ast}(
u |x)|x)=Q^{-1}(Q^{\ast}( u |x)|x)$ yields $Q^{\ast}( u |x)=Q( u
|x)$.

Claim (5).  We have $Y_x=Q(U|x)$ has quantile function $Q^{\ast}$. A
quantile function is known to be equivariant to monotone increasing
transformations, including location-scale transformations. Thus,
this is true in particular for $Q^{\ast}$.

Claim (6) is immediate from claim (3).

Claim (7). The proof of continuity of $F$ is subsumed in the step 1
of the proof of Proposition 3 (see below).  Therefore, for any
sequence $x_{t} \rightarrow x$ we have that $F(y|x_t) \to F(y|x)$
uniformly in $y$, and $F$ is continuous. Let $u_t \to u$ and $x_t
\to x$. Since $F(y|x) = u$ has a unique root $y= Q^{\ast}(u|x)$, the
root of $F(y|x_t) = u_t$, i.e., $y_t =Q^{\ast}(u_t|x_t)$, converges
to $y$ by a standard argument, see, e.g., van der Vaart and Wellner
(1997). $\square $


\subsection{Proof of Propositions \ref{proposition: hadamard_diff}--\ref{proposition: limit_distribution_linear_func}}

In the proofs that follow we will repeatedly use Lemma 1, which
establishes the equivalence of continuous convergence and uniform
convergence:

\begin{lemma} Let $\mathbb{D}$ and $\mathbb{D}'$ be complete separable metric spaces,
with $\mathbb{D}$ compact. Suppose $f: \mathbb{D} \mapsto
\mathbb{D}'$ is continuous. Then a sequence of functions $f_n:
\mathbb{D} \mapsto \mathbb{D}'$ converges to $f$ uniformly on
$\mathbb{D}$ if and only if for any convergent sequence $x_n \to x$
in $\mathbb{D}$ we have that $f_n(x_n) \to f(x)$.

\end{lemma}

\noindent \textbf{Proof of Lemma 1:} See, for example, Resnick
(1987), page 2. $\square $

\textbf{Proof of Proposition \ref{proposition: hadamard_diff}.}

\textbf{Part 1.} We have that for any $\delta >0$, there exists $\epsilon >0$ such that for $%
 u \in B_{\epsilon }( u _{k}(y|x))$ and for small enough $t\geq 0$
\begin{equation}
1\{Q(u|x )+th_t( u |x)\leq y\}\leq 1\{Q(u|x )+t(h( u
_{k}(y|x)|x)-\delta )\leq y\},  \notag
\end{equation}%
for all $k\in \{1,2, ...,K(y|x)\}$; whereas for all $ u \not\in \cup
_{k}B_{\epsilon }( u _{k}(y|x))$, as $t \rightarrow 0$,
\begin{equation}
1\{Q(u|x )+th_t( u |x)\leq y\}=1\{Q(u|x )\leq y\}.  \notag
\end{equation}%
Therefore,
\begin{eqnarray}
&&\frac{\int_{0}^{1}1\{Q(u|x )+th_t( u |x)\leq y\}d u
-\int_{0}^{1}1\{Q( u|x )\leq y\}d u }{t}  \label{epression1} \\
&\leq &\sum_{k=1}^{K(y|x)}\int_{B_{\epsilon }( u _{k}(y|x))}\frac{%
1\{Q(u|x )+t(h( u _{k}(y|x)|x)-\delta)\leq y\}-1\{Q(u|x )\leq y\}}{t%
}d u,  \notag
\end{eqnarray}%
which by the change of variable $y' = Q(u|x)$ is equal to
\begin{equation*}
\frac{1}{t}\sum_{k=1}^{K(y|x)} \int_{J_k \cap [y, y -t(h( u
_{k}(y|x)|x)-\delta)]} \frac{1}{|\partial_u Q(Q^{-1}(y'|x)|x)|} dy',
\end{equation*}%
where $J_k$ is the image of $B_{\epsilon }( u _{k}(y|x))$ under
$u\mapsto Q( \cdot |x)$.  The change of variable is possible because
for $\epsilon$ small enough, $Q(\cdot|x)$ is one-to-one between
$B_{\epsilon }( u _{k}(y|x))$ and $J_k$.

Fixing $\epsilon>0$, for $t \to 0$, we have that $J_k \cap [y, y
-t(h( u _{k}(y|x)|x) - \delta)] = [y, y -t(h( u
_{k}(y|x)|x)-\delta)],$ and $|\partial_u Q(Q^{-1}(y'|x)|x)| \to
|\partial_u Q(u _{k}(y|x)|x)|$ as $Q^{-1}(y'|x) \to u _{k}(y|x)$.
Therefore, the right hand term in (\ref{epression1}) is no greater
than%
\begin{equation}
\sum_{k=1}^{K(y|x)}\frac{-h( u _{k}(y|x)|x)+\delta}{|\partial_u Q( u
_{k}(y|x)|x)|}+o\left( 1\right) .  \notag
\end{equation}
Similarly $ \sum_{k=1}^{K(y|x)}\frac{-h( u
_{k}(y|x)|x)-\delta}{|\partial_u Q( u _{k}(y|x)|x)|}+o\left(
1\right) $ bounds (\ref{epression1}) from below. Since $\delta >0$
can be made arbitrarily small, the result follows.

To show that the result holds uniformly in $\left( y, x \right) \in
K$, a compact subset of $\mathcal{YX}^{\ast }$, we use Lemma 1. Take
a sequence of $(y_t, x_t)$ in $K$ that converges to $(y,x)\in K$,
then the preceding argument applies to this sequence, since (1) the
function $(y,x) \mapsto -h( u _{k}(y|x)|x)/|\partial_u Q( u
_{k}(y|x)|x)|$ is uniformly continuous on $K$, and (2) the function
$(y,x) \mapsto K(y|x)$ is
 uniformly continuous on $K$.  To see (2), note that $K$ excludes a
neighborhood of critical points
$(\mathcal{Y}\setminus\mathcal{Y}^*_x, x\in \mathcal{X})$, and
therefore can be expressed as the union of a finite number of
compact sets $(K_1,...,K_M)$ such that the function $K(y|x)$ is
constant over each of these sets, i.e., $K(y|x) = k_j$ for some
integer $k_j>0$, for all $(y,x) \in K_j$ and $j \in \{1, ...,M\}$.
Likewise, (1) follows by noting that the limit expression for the
derivative is continuous on each of the sets $(K_1,...,K_M)$ by the
assumed continuity of $h(u|x)$ in both arguments, continuity of
$u_k(y|x)$ (implied by the Implicit Function Theorem), and the
assumed continuity of $\partial_u Q(u|x)$ in both arguments.
$\square $

\textbf{Part 2}. For a fixed $x$ the result follows by Part 1 of
Proposition \ref{proposition: hadamard_diff}, by step 1 of the proof
below, and by an application of the Hadamard differentiability of
the quantile operator shown by Doss and Gill (1992). Step 2
establishes uniformity over $x\in \mathcal{X}$.

Step 1. Let $K$ be a compact subset of $\mathcal{YX}^*$. Let
$(y_t,x_t)$ be a sequence in $K$, convergent to a point, say
$(y,x)$. Then, for every such sequence, $ \epsilon_t := t
\|h_t\|_{\infty} + \| Q(\cdot|x_t) -Q (\cdot|x)\|_{\infty} + |y_t
-y|  \to 0$, and
 \begin{eqnarray}  \notag
|F(y_t|x_t,h_{t}) - F(y|x)| & \leq & \Big |\int_{0}^{1} [1\{Q (u|x_t
)+th_t( u |x)\leq y_t\} - 1\{Q(u|x )\leq y\}]d u \Big | \\
 &
\leq & \Big |\int_{0}^{1}1\{ |Q(u|x) -y| \leq \epsilon_t \} du \Big
| \to 0,
 \end{eqnarray}
where the last step follows from the absolute continuity of $y
\mapsto F(y|x)$, the distribution function of $Q(U|x)$. By setting
$h_t=0$ the above argument also verifies that  $F(y|x)$ is
continuous in $(y,x)$. Lemma 1 implies uniform convergence of
$F(y|x,h_t)$ to $F(y|x)$, which in turn implies by a standard
argument\footnote{See, e.g., Lemma 1 in  Chernozhukov and
Fernandez-Val (2005).} the uniform convergence of quantiles
$Q^{\ast}( u |x,h_{t})\to Q^{\ast}( u |x)$, uniformly over $K^*$,
where $K^*$ is any compact subset of $\mathcal{UX^*}$.

 Step 2. We have that uniformly over $K^*$,
\begin{equation}
\begin{split}\label{Att}
\frac{F(Q^{\ast}( u |x,h_{t})|x,h_{t})-F(Q^{\ast}( u
|x,h_{t})|x)}{t}&
=D_{h}(Q^{\ast}( u |x,h_{t})|x)+o(1), \\
& =D_{h}(Q^{\ast}( u |x)|x)+o(1),
\end{split}%
\end{equation}%
using Step 1, Proposition 2,  and the continuity properties of
$D_{h}(y|x)$. Further, uniformly over $K^*$, by Taylor expansion and
Proposition 1, as $t \rightarrow 0$,
\begin{equation}
\frac{F(Q^{\ast}( u |x,h_{t})|x)-F(Q^{\ast}( u |x)|x)}{t} =
f(Q^{\ast}( u |x)|x)\frac{Q^{\ast}( u |x,h_{t})-Q^{\ast}( u |x)}{t}
+ o(1), \label{Bt}
\end{equation}%
and (as will be shown below)
\begin{equation}
\frac{F(Q^{\ast}( u |x,h_{t})|x,h_{t})-F(Q^{\ast}( u
|x)|x)}{t}=o(1), \label{Ct}
\end{equation}%
as $t \rightarrow 0$. Observe that the left hand side of (\ref{Ct})
equals that of (\ref{Bt}) plus that of (\ref{Att}). The result then
follows.

It only remains to show that equation (\ref{Ct}) holds uniformly in
$K^*$. Note that for any right-continuous cdf $F$, we have that $ u
\leq F(Q^{\ast}( u ))\leq
 u +F(Q^{\ast}( u ))-F(Q^{\ast}( u )-)$, where $F(\cdot -)$ denotes the left
 limit of $F$, i.e., $F(x_{0}-) = \lim_{x \uparrow x_{0}} F(x)$. For any continuous,
strictly increasing cdf $F$, we have that $F(Q^{\ast}( u ))= u $.
Therefore, write
\begin{equation*}
\begin{split}
& 0\leq \frac{F(Q^{\ast}( u |x,h_{t})|x,h_{t})-F(Q^{\ast}( u |x)|x)}{t} \\
& \quad \leq \frac{ u +F(Q^{\ast}( u |x,h_{t})|x,h_{t})-F(Q^{\ast}(
u
|x,h_{t})-|x,h_{t})- u }{t} \\
&  \quad \leq \frac{F(Q^{\ast}( u |x,h_{t})|x,h_{t})-F(Q^{\ast}( u
|x,h_{t})-|x,h_{t}) }{t} \\
 & \quad \overset{(1)}=\frac{[F(Q^{\ast}( u |x,h_{t})|x,h_{t})-F(Q^{\ast}( u
|x,h_{t})|x)]}{t}\\
&  \quad  - \frac{[F(Q^{\ast}( u |x,h_{t})-|x,h_{t})-F(Q^{\ast}( u |x,h_{t})-|x)]}{t} \\
& \quad \overset{(2)}=D_{h}(Q^{\ast}( u |x,h_{t})|x)-D_{h}(Q^{\ast}(
u |x,h_{t})-|x)+o(1)=o(1),
\end{split}
\end{equation*}
as $t \rightarrow 0$, where in (1) we use that $F(Q^{\ast}( u
|x,h_{t})|x) = F(Q^{\ast}( u |x,h_{t})-|x)$ since $F(y|x)$ is
continuous and strictly increasing in $y$, and in (2) we use
Proposition 2. $\square $

The following lemma, due to Pratt (1960), will be very useful to
prove Proposition 4.

\begin{lemma} Let  $|f_n| \leq G_n$ and suppose that $f_n \to f$
and $G_n \to G$ almost everywhere, then if $\int G_n \to \int G$
finite,  then $\int f_n \to \int f$.
\end{lemma}

\textbf{Proof of Lemma 2.} See Pratt (1960). $\square$  


\begin{lemma}[Boundedness and Integrability Properties]
Under the hypotheses of Proposition \ref{proposition:
hadamard_diff},  we  have that, for all $(u,x) \in \mathcal{UX}$,
\begin{eqnarray}\label{mmm0}
 & & |\widetilde D_{h_t}(u|x,t)| \leq \|h_t\|_{\infty},
\end{eqnarray}
and, for all $(y,x) \in \mathcal{YX}$,
\begin{eqnarray} \label{mmm1}
 & & |D_{h_t}(y|x,t)| \leq \Delta(y|x,t) = \int_0^1  \frac{1\{ |Q(u|x) -y| \leq t
 \|h_t\|_{\infty}
 \}}{t} du,
\end{eqnarray}
where for any $x_t \to x \in \mathcal{X}$, as $t \rightarrow 0$,
$$ \text{ $\Delta(y|x_t,t) \to 2 \|h\|_{\infty} f(y|x) $ for a.e $y
\in \mathcal{Y}$  and $ \int_{\mathcal{Y}} \Delta(y|x_t,t) dy \to
\int_{\mathcal{Y}} 2 \|h\|_{\infty} f(y|x) dy $. }
$$
\end{lemma}

\textbf{Proof of Lemma 3.}  To show (\ref{mmm0}) note that
\begin{eqnarray}
 & & \sup_{(u,x) \in \mathcal{UX}} |\widetilde D_{h_t}(u|x,t)| \leq \|h_t\|_{\infty}
\end{eqnarray}
immediately follows from the equivariance property noted in Claim
(5) of Proposition 1.

The inequality (\ref{mmm1}) is trivial.  That for any $x_t \to x \in
\mathcal{X}$, $\Delta(y|x_t,t) \to 2 \|h\|_{\infty} f(y|x) $ for a.e
$y \in \mathcal{Y}$ follows by applying Proposition 2 respectively
with functions $h'_t(u|x)= \|h_t\|_{\infty}$ and $h'_t(u|x) =
-\|h_t\|_{\infty}$ (for the case when $f(y|x)>0$; and trivially
otherwise). Similarly, that for any $y_t \to y \in \mathcal{Y}$, $
\text{ $\Delta(y_t|x,t) \to 2 \|h\|_{\infty} f(y|x) $ for a.e $x \in
\mathcal{X}$ } $ follows by Proposition 2 (for the case when
$f(y|x)>0$; and trivially otherwise) .

 Further, by Fubini's Theorem,
\begin{eqnarray}\label{mmm2}
\int_{\mathcal{Y}} \Delta(y|x_t,t) dy & = & \int_0^1
 \underbrace{\left ( \int_{\mathcal{Y}} \frac{1\{ |Q(u|x_t) -y| \leq t
 \|h_t\|_{\infty}
 \}}{t} dy \right)}_{=: \ f_t(u)} du.
 \end{eqnarray}
Note that $f_t(u) \leq 2\|h_t\|_\infty$. Moreover, for almost every
$u$,  $f_t(u) = 2\|h_t\|_\infty$ for small enough $t$, and
$2\|h_t\|_\infty$ converges to $2\|h\|_\infty$ as $t \to 0$. Then,
trivially, $ 2\int_0^1 \|h_t\|_{\infty} du \to 2\|h\|_{\infty}$. By
 Lemma 2  the right hand side of (\ref{mmm2})
converges to $2\|h\|_{\infty}$. $\square$

\subsection{Proof of Proposition \ref{proposition: hadamard_diff linear_funct}} Define  $m_t(y|x,y'):=
g(y|x,y') D_{h_t}(y|x,t)$ and $m(y|x,y') := g(y|x,y') D_{h}(y|x)$.
To show claim (1),  we need to demonstrate that for any $y'_t \to
y'$ and $x_t \to x$
\begin{eqnarray}\label{claim1}
\int_{\mathcal{Y}} m_t(y|x_t,y'_t) d y \to \int_{\mathcal{Y}}
m(y|x,y') d y,
 \end{eqnarray}
and that the limit is continuous in $(x,y')$.  We have that
$|m_t(y|x_t,y_t)|$ is bounded, for some constant $C$, by $ C
\Delta(y|x_t,t) $ which converges a.e. and the integral of which
converges to a finite number by Lemma 3.  Moreover, by Proposition
2, for almost every $y$ we have $ m_t(y|x_t,y_t') \to m(y|x,y').$ We
conclude that (\ref{claim1}) holds by Lemma 2.

In order to check continuity, we need to show that for any $y'_t \to
y'$ and $x_t \to x$
\begin{eqnarray}\label{cont}
\int_{\mathcal{Y}} m (y|x_t,y'_t) d y \to \int_{\mathcal{Y}}
m(y|x,y') d y.
 \end{eqnarray}
 We have that $m(y|x_t,y'_t) \to m(y|x,y')$ for almost every $y$.
 Moreover,  $m(y|x_t,y_t)$ is dominated by $2\|g\|_{\infty} \|h\|_{\infty}
 f(y|x_t)$, which converges to $2\|g\|_{\infty} \|h\|_{\infty}
 f(y|x)$ for almost every $y$, and, moreover, $\int_{\mathcal{Y}}\|g\|_{\infty} \|h\|_{\infty}
 f(y|x)dy$ converges to $\|g\|_{\infty} \|h\|_{\infty}$. Conclude that (\ref{cont}) holds by Lemma 2.




To show claim (2), define  $m_t(u|x,u') = g(u|x,u')
\tilde D_{h_t}(u|x)$ and $m(u|x,u') = g(u|x,u') \tilde
D_{h}(u|x)$.  Here we need to show that for any $u'_t
\to u'$ and $x_t \to x$
\begin{eqnarray}\label{claim3}
\int_{\mathcal{U}} m_t(u|x_t,u'_t) d u \to \int_{\mathcal{U}}
m(u|x,u') d u,
 \end{eqnarray}
and that the limit is continuous in $(u',x)$. We have that
$m_t(u|x_t,u'_t)$ is bounded by $ g(u|x_t) \|h_t\|_{\infty}$, which
converges to $ g(u|x) \|h\|_{\infty}$ for a.e. $u$. Furthermore, the
integral of $ g(u|x_t) \|h_t\|_{\infty}$ converges to the integral
of $g(u|x) \|h\|_{\infty}$ by the dominated convergence theorem.
Moreover, by Proposition 2, we have that $ m_t(u|x_t,u'_t) \to
m(u|x,u')$  for almost every $u$. We conclude that (\ref{claim3})
holds by Lemma 2.

In order to check the continuity of the limit, we need to show that
for any $u'_t \to u'$ and $x_t \to x$
\begin{eqnarray}\label{cont2}
\int_{\mathcal{U}} m (u|x_t,u_t') d u \to \int_{\mathcal{U}}
m(u|x,u') d u.
 \end{eqnarray}
 We have that $m(u|x_t,u'_t) \to m(u|x,u')$ for almost every $u$.
 Moreover, for small enough $t$, $m(u|x_t,u_t')$ is dominated by
 $|g(u|x_t, u_t')| \|h\|_{\infty}$, which converges for almost every value of $u$
to  $|g(u|x, u')| \|h\|_{\infty}$ as $t \to 0$.
Furthermore, the integral of $|g(u|x_t, u_t')|
\|h\|_{\infty}$ converges to the integral of $|g(u|x,
u')| \|h\|_{\infty}$ by the dominated convergence
theorem. We conclude that (\ref{cont2}) holds by Lemma
2. $\square$

\subsection{Proof of Proposition \ref{proposition: limit_distribution}}  This proposition simply follows by the
functional delta method (e.g., van der Vaart, 1998). Instead of
restating what this method is, it takes less space to simply recall
the proof in the current context.

To show the first part, consider the map $
g_{n}(y,x|h)=a_n(F(y|x,h/a_n)-F(y|x))$. The sequence of maps
satisfies $g_{n^{\prime }}(y,x|h_{n^{\prime }})\rightarrow
D_{h}(y|x)$ in $\ell^{\infty}(K)$ for every subsequence
$h_{n^{\prime }}\rightarrow h$ in $\ell ^{\infty }(\mathcal{UX})$,
where $h$ is continuous. It follows by the extended continuous
mapping theorem that, in $\ell^{\infty}(K)$,
$g_{n}(y,x|a_n(\widehat{Q}( u |x)-Q( u |x)))\Rightarrow D_{G}(y|x)$
as a stochastic process indexed by $(y,x)$, since $a_n(\widehat{Q}(
u |x)-Q( u |x)) \Rightarrow G(u|x)$ in
$\ell^{\infty}(\mathcal{UX})$.

Conclude similarly for the second part. $\square $

\subsection{Proof of Proposition \ref{proposition: limit_distribution_linear_func}}  This follows by the
functional delta method, similarly to the proof of Proposition
\ref{proposition: limit_distribution}. $\square $

\end{document}